%% file: main.tex
\documentclass[acmlarge, imwut]{acmart}

\usepackage[export]{adjustbox}
\usepackage{enumitem}
\usepackage{multirow}
\usepackage{arydshln}
\PassOptionsToPackage{table}{xcolor}
\usepackage{colortbl}
\usepackage[capitalise]{cleveref}

\usepackage{graphicx}

\AtBeginDocument{%
  }

\setcopyright{acmlicensed}
\copyrightyear{2026}
\acmYear{2026}
\acmDOI{}

\begin{document}

\title{DISCOVER: Identifying Patterns of Daily Living in Human Activities from Smart Home Data}

\author{Alexander Karpekov}
\email{alex.karpekov@gatech.edu}
\orcid{0009-0006-5317-2991}
\affiliation{%
  \institution{Georgia Institute of Technology}
  \city{Atlanta}
  \state{Georgia}
  \country{USA}
}

\author{Archith Iyer}
\email{aiyer351@gatech.edu}
\orcid{0009-0002-3559-6038}
\affiliation{%
  \institution{Georgia Institute of Technology}
  \city{Atlanta}
  \state{Georgia}
  \country{USA}
}

\author{Sourish Gunesh Dhekane}
\email{sourish.dhekane@gatech.edu}
\orcid{0009-0004-0035-819X}
\affiliation{%
  \institution{Georgia Institute of Technology}
  \city{Atlanta}
  \state{Georgia}
  \country{USA}
}

\author{Sonia Chernova}
\email{chernova@gatech.edu}
\orcid{0000-0001-6320-0825}
\affiliation{%
  \institution{Georgia Institute of Technology}
  \city{Atlanta}
  \state{Georgia}
  \country{USA}
}

\author{Thomas Pl{\"o}tz}
\email{thomas.ploetz@gatech.edu}
\orcid{0000-0002-1243-7563}
\affiliation{%
  \institution{Georgia Institute of Technology}
  \city{Atlanta}
  \state{Georgia}
  \country{USA}
}


\renewcommand{\shortauthors}{Karpekov et al.}

\newcommand{\ToolName}{DISCOVER}

\begin{abstract}
\input{sections/0_abstract}
\end{abstract}



\maketitle

 \thispagestyle{fancy}
\fancyhead{} 
\fancyfoot{} 
\pagenumbering{gobble}
\fancyfoot[C]{\textcolor{red}{This manuscript is under review. Please contact alex.karpekov@gatech.edu for up-to-date information}}

\input{sections/1_intro}

\input{sections/2_related_work}

\input{sections/3_approach}

\input{sections/4_data}

\input{sections/5_labeling_quality}

\input{sections/6_supervised_benchmark}

\input{sections/7_discussion}

\input{sections/8_conclusion}

\bibliographystyle{ACM-Reference-Format}
\bibliography{ref}

\newpage
\input{sections/9_appendix}

\end{document}

%% file: sections/0_abstract.tex
\noindent
Smart homes equipped with ambient sensors offer a transformative approach to continuous health monitoring and assisted living. Traditional research in this domain primarily focuses on Human Activity Recognition (HAR), which relies on mapping sensor data to a closed set of predefined activity labels. However, the fixed granularity of these labels often constrains their practical utility, failing to capture the subtle, household-specific nuances essential, for example, for tracking individual health over time. To address this, we propose \ToolName{}, a framework for \textit{discovering} and annotating \underline{P}atterns of \underline{D}aily \underline{L}iving (PDL)---fine-grained, recurring sequences of sensor events that emerge directly from a resident's unique routines. \ToolName{} utilizes a self-supervised feature extraction and representation-aware clustering pipeline, supported by a custom visualization interface that enables experts to interpret and label discovered patterns with minimal effort. Our evaluation across multiple smart-home environments demonstrates that \ToolName{} identifies cohesive behavioral clusters with high inter-rater agreement while achieving classification performance comparable to fully-supervised baselines using only $0.01\%$ of the labels. Beyond reducing annotation overhead, \ToolName{} establishes a foundation for longitudinal analysis. By grounding behavior in a resident’s specific environment rather than rigid semantic categories, our framework facilitates the observation of within-person habitual drift. This capability positions the system as a potential tool for identifying subtle behavioral indicators associated with early-stage cognitive decline in future longitudinal studies.

%% file: sections/1_intro.tex
\section{Introduction}

\textit{What is the precise definition of a human activity in the context of smart homes?} Currently, no consensus exists in the literature, primarily because the relevance and scope of an activity are inherently dependent on the specific application \cite{dhekane2025transfer}. A large portion of research studying human activities in smart homes views it as a human activity recognition (HAR) problem \cite{ahmed2022clustering, deepcasas2018, casas2009}. HAR promises to provide a foundation for longitudinal assessments of changes in daily habits, which  can facilitate the early detection of emerging physical or cognitive challenges, enabling timely interventions and improved patient outcomes \cite{morita2023health, riboni2016analysis}.
In this setting, HAR relies on a fixed vocabulary of activity labels, such as ``Cooking'', ``Bathing'', and ``Sleeping''. 
These specific activities are usually derived from Activities of Daily Living (ADL) -- the basic skills necessary for individuals to independently care for themselves \cite{edemekong2025adl}. However, day-to-day behavioral patterns in a given home do not neatly conform to these standardized definitions, frequently involving personal habits and unique household routines that fall outside the scope of general-purpose activity datasets.
For example, the activity of ``Relaxing'' across various smart homes can exhibit a wide range of different physical manifestations, such as strolling around the house, listening to music while sitting in a chair, or watching TV.  
Furthermore, many daily activities do not correspond to any of the predefined labels, resulting in large catch-all labels that provide no useful insights into the underlying behavior of residents.

An alternative to the traditional activity-centric HAR paradigm is the identification of \underline{P}atterns of \underline{D}aily \underline{L}iving (PDL). PDLs represent meaningful, fine-grained, and recurring sequences of sensor events that capture the data-driven structure of how residents interact with their specific environments. Rather than forcing human behavior into a limited set of predefined categories, PDLs emerge from latent regularities in sensor activations across space and time. 
These patterns may encompass canonical ADLs such as ``Cooking'', but they also capture the finer interactions that constitute them, such as ``Stove use'' or ``Table setting.'' Furthermore, PDLs reveal transitional behaviors (e.g., movement from bedroom to bathroom) and highly personalized routines (such as the specific use of a particular armchair) that lack a distinct semantic equivalent in traditional activity taxonomies.
By uncovering these patterns directly from the data, smart home systems can reveal the idiosyncratic organization of daily life in unique environments and, crucially, quantify how these routines evolve or deviate over time.

This distinction between predefined labels and PDLs carries significant practical implications for both modeling and practical interpretation. Traditional activity-centric approaches prioritize alignment with a closed set of annotations and are typically evaluated using standard classification metrics such as accuracy and F1 score. However, these metrics offer little insight into the underlying structure of those activities or the unique, user-specific habits manifested within a given environment. In contrast, pattern-centric approaches emphasize uncovering latent structures within continuous, unsegmented sensor data, even when those structures subdivide or intersect with traditional activity labels. While PDLs can be mapped back to canonical ADLs for benchmarking, their primary value lies in providing a flexible, high-resolution representation of daily life, which can facilitate personalized monitoring and longitudinal sensitivity analysis.

In this work, we build upon this pattern-centric viewpoint by proposing \ToolName{}, a framework designed to \textit{discover} PDLs from smart home data without requiring prior annotations or pre-segmentation. \ToolName{} identifies these patterns through self-supervised pre-training with a masked modeling objective, generating semantically rich sensor data embeddings. These representations are further refined using a clustering objective to produce distinct clusters that correspond to unique Patterns of Daily Living. To facilitate the interpretation of these clusters, we provide a user-friendly visualization tool that replays sensor activation sequences directly onto a smart home layout, enabling a seamless labeling process accessible to non-experts. By annotating as little as $0.01\%$ of the data samples, \ToolName{} propagates labels across all discovered patterns, streamlining the entire annotation process. Crucially, by removing the need for manual pre-segmentation, our approach allows continuous sensor streams to organize naturally according to their inherent behavioral structure while preserving both interpretability and user control.

The main contributions of our work are as follows:
\begin{enumerate}
    \item \textbf{Data-Grounded Pattern Discovery}: We develop a self-supervised feature extraction and clustering pipeline that learns directly from raw, continuous sensor streams without the need to rely on large annotated training datasets to uncover data-driven patterns of daily living. 
    Rather than fitting the data to a closed set of predefined activities, the system \textit{discovers} environment-specific behavioral structures grounded in the sensor data itself.

    \item \textbf{Eliminating Pre-Segmentation Constraints}: 
    By handling continuous, unsegmented time-series, our framework is applicable to realistic smart home scenarios and can be easily extended to real-world deployments. 

    \item \textbf{Efficient Human-in-the-Loop Annotation}: Leveraging an active-learning-like strategy, we focus manual annotation effort on a sparse set of representative cluster centroids. By propagating labels from a small set of representative cluster centroids, \ToolName{} significantly reduces the manual annotation burden. In our experiments, we achieved high classification performance by annotating only a handful of examples per cluster, representing less than $0.01\%$ of the total dataset.

    \item \textbf{Open-Source PDL Visualization Tool}: We introduce an interactive, open-source visualization tool that enables researchers and practitioners to explore \textit{discovered} Patterns of Daily Living, inspect their spatial and temporal structure, and relate them back to sensor layouts and canonical activities. 
    The tool supports qualitative interpretation, providing a practical bridge between unsupervised pattern \textit{discovery} and validation.

\end{enumerate}

To validate our approach, we apply \ToolName{} to a widely used smart-home dataset spanning multiple smart home environments. Our analysis demonstrates that \ToolName{} yields coherent behavioral clusters that align with canonical activities while uncovering fine-grained structures beyond current annotation schemas. We demonstrate the practical value of this framework by showing that it achieves classification performance comparable to fully-supervised baselines with orders of magnitude fewer labels, while simultaneously enabling the detection of subtle, longitudinal habitual shifts. Crucially, our findings indicate that these discovered patterns provide a higher-resolution representation of daily life than traditional HAR labels. By anchoring behavior in a resident’s specific environment, \ToolName{} bridges the gap between raw sensor data and the nuanced longitudinal insights required for future research into early-stage cognitive decline.

%% file: sections/2_related_work.tex
\section{Background \& Related Work}
\label{sec:related_work}

\ToolName{} introduces a novel perspective for analyzing smart home data through the discovery of Patterns of Daily Living (PDL). This approach looks beyond traditional ground truth annotations, which typically relate to only a limited subset of activity granularities and often obscure the underlying behavioral motifs. By shifting the focus from predefined labels to data-grounded \textit{discovery}, we establish a more flexible framework for understanding longitudinal habits.

We structure our review of related work as follows. First, \cref{subsec:act_levels} defines the hierarchical spectrum of activity granularity. We then survey benchmark smart home datasets through the lens of sensing density and its impact on the activity taxonomy in \cref{sec:related:sensing}. In \cref{subsec:viz_annotate_tools}, we evaluate existing visualization and annotation methods, highlighting the reliability gaps in standard labeling protocols. \cref{subsec:har_pipeline} provides an overview of the traditional smart home HAR pipeline and its current limitations. Finally, \cref{subsec:clustering_sh} reviews the state-of-the-art in self-supervised feature extraction and clustering, establishing the technical foundation for the \ToolName{} framework.

\subsection{Granularity Levels of Human Activities}
\label{subsec:act_levels}

Human activity is not a monolithic concept but rather a hierarchical spectrum ranging from primitive physical movements to complex, longitudinal behaviors and habits \cite{bobick1997movement, nagel1988image, vrigkas2015review}. At the most granular level, behaviors are composed of movements or gestures, comprised of primitive, atomic, and recurring motion units. For example, the sequential movements of lifting a leg, taking a stride, and pushing the ground backward constitute the more complex event of taking a step. When performed within a specific context, these event sequences generate activities (e.g., ``Walking''). Finally, a high-level aggregation of activities denotes a behavior or a habit, a more abstract construct that typically requires longitudinal analysis to identify.

This taxonomy has been extensively studied in computer vision, where capturing micro-level gestures and environmental context from video is relatively straightforward \cite{bukht2025review, bobick1997movement, nagel1988image, arshad2022human}. 
However, achieving a similar resolution using sparse modalities, 
such as ambient smart-home sensors, 
remains a significant challenge. 
While existing work has explored identifying activity patterns in smart environments \cite{boovaraghavan2023tao, bourobou2015user, alia2025identifying, yassine2017mining, ni2015elderly}, the interpretation of these patterns varies wildly, ranging from high-level behavioral trends to mid-level actions.

In contrast, \ToolName{} assumes having access to only ambient sensor data and is designed to operate specifically at the fine-grained pattern level, identifying the recurring motifs that bridge the gap between raw sensor triggers and semantic activities, without using any ground truth activity annotations. 
We distinguish \ToolName{} from traditional approaches by three core guarantees. First, the \textit{discovery} is grounded in the actual data: patterns emerge directly from latent regularities in the sensor stream rather than conforming to predefined, top-down categories. Second, each discovered cluster represents a single, semantically consistent routine (e.g., a specific path between the dining table and the kitchen). Third, these patterns are identified without any prior labels, bypassing the primary bottleneck of supervised HAR.

\subsection{Sensing Density and Activity Taxonomy}
\label{sec:related:sensing}

The granularity of activity labels in smart home research is fundamentally constrained by the density and modality of the underlying sensor deployment. As established in \cref{subsec:act_levels}, human behavior exists on a hierarchical spectrum of movements, actions, activities, and behaviors. Existing smart home datasets typically occupy specific tiers of this hierarchy based on the instrumentation of their respective deployments.

Most widely-used benchmarks, such as CASAS \cite{cook2012casas}, Van Kasteren \cite{van2010activity}, and ARAS \cite{alemdar2013aras}, represent sparse ambient sensing environments. These deployments typically utilize motion and contact sensors limited to one to two per room or primary appliance \cite{sensorDatasetSurvey}. Consequently, these datasets are primarily annotated at the activity level, capturing canonical ADLs like Cooking, Sleeping, or Personal Hygiene. While these labels are semantically clear, they often collapse heterogeneous behaviors into single, non-descriptive categories. For example, a Relax label may encompass reading in an armchair, watching television, or napping. The sparse nature of these deployments inherently limits the tracking of the finer-grained actions required to distinguish these sub-routines.

Conversely, datasets like Placelab \cite{intille2005placelab} and the MIT activity recognition datasets \cite{tapia2004activity} utilize significantly denser instrumentation, often including hundreds of sensors to monitor fine-grained interactions. These datasets provide annotations at the action or movement granularity. While these environments offer a rich view of human-object interaction, such extensive instrumentation is difficult to scale to real-world residential settings due to the prohibitive complexity and maintenance of the hardware \cite{boovaraghavan2023tao}.

Rather than being restricted by a specific sensing modality, \ToolName{} is designed to be configuration-adaptive. It does not rely on pre-defined activity mappings that may not be supported by the available instrumentation. Instead, \ToolName{} decomposes available data into meaningful Patterns of Daily Living (PDLs). \ToolName{} is capable of identifying localized sub-routines, such as specific trajectories through a kitchen or stationary seating preferences, that standard HAR pipelines typically discard as non-descriptive noise. This approach allows \ToolName{} to extract high-resolution behavioral insights from existing sparse datasets without requiring the dense instrumentation of lab-based deployments.

\subsection{Smart Home Visualization and Annotation Methods}
\label{subsec:viz_annotate_tools}

The hierarchical temporal patterns inherent in smart home data require robust visualization and annotation protocols. While specific actions, such as taking medicine, can be resolved via simple pillbox sensors, broader activities like Cooking often comprise a stochastic sequence of sub-activities (e.g., pantry access, food preparation, and stove usage) captured only through ambient motion triggers. Furthermore, the temporal disparity between long-duration events, such as sleeping, and near-instantaneous transitions, such as leaving home, complicates the task of consistent data interpretation. 

Visualization in the current literature is frequently relegated to ``displaying insights'' rather than facilitating raw data exploration \cite{castelli2017happened, le2014design}. Existing tools for the CASAS datasets primarily utilize spatial video overlays \cite{karpekov2025visar} and heatmaps \cite{chen2011casasviz} to provide context for temporal sensor activation patterns. However, the annotation protocols associated with these visualizations remain a significant bottleneck. Standard methods often rely on scheduled routines \cite{marble2022, orange2017}, which lack the naturalistic, ``in-the-wild'' component, or retrospective self-reporting \cite{alemdar2013aras}, which is notoriously prone to recall bias and high participant  \cite{tonkin2018talk}. Alternatively, manual post-hoc labeling of sensor logs, as performed in CASAS, is labor-intensive and susceptible to systematic human error and inter-rater \cite{demrozi2023comprehensive}. 

These annotation inaccuracies can fundamentally undermine the reliability of HAR benchmarks and subsequent model evaluation. \ToolName{} addresses these deficiencies by automating the identification of Patterns of Daily Living (PDLs), requiring expert validation for only a minimal subset of the data. By integrating an intuitive spatial visualization via our custom-made tool \cite{karpekov2025visar}, \ToolName{} allows non-experts to accurately interpret and label discovered patterns, thereby establishing a scalable and more reliable ground truth than traditional manual protocols.

\subsection{Activity Recognition in Smart Homes}
\label{subsec:har_pipeline}

Similar to the case with body-worn IMU sensors \cite{bulling2014tutorial}, human activity recognition (HAR) in smart homes is based on a multi-stage processing pipeline consisting of data acquisition, pre-processing, segmentation, feature extraction, and classification.   
HAR in smart environments has emerged as a critical area of research, enabling the identification of Activities of Daily Living (ADLs) through ambient sensors \cite{bouchabou2021survey, vanKasteren2010, Rafferty2017, kientz2008georgia}. ADLs are particularly valuable for routine detection, which feeds numerous downstream applications, including health monitoring and assistive systems for independent living \cite{chatting2023automated, csurka2017domain, morita2023health}. Conventional approaches to HAR rely  on supervised learning methods, where large amounts of labeled data are needed. These methods span a range of classic machine-learning techniques, including Random Forests \cite{sedky2018evaluating}, Hidden Markov Models (HMMs) and Support Vector Machines (SVMs) \cite{cook2010learning}. A majority of these approaches rely on features that are hand-crafted by domain experts, and universally assume pre-segmented data, which limits their applicability in real-world scenarios.

Advances in deep learning have expanded the capabilities of HAR systems. Deep Neural Networks (DNNs) \cite{alaghbari2022activities}, Long Short-Term Memory networks ((LSTM) \cite{liciotti2020sequential}, and hybrid architectures with Embeddings from Language Models (ELMo) combined with BiLSTM
\cite{bouchabou2021using} have demonstrated promising
performance in activity classification tasks. Additionally, graph neural networks (GNNs) have been employed to capture spatial and temporal relationships in sensor data \cite{li2019relation, zhou2020graph, plotz2023know}. 

Transfer learning is another promising subdomain, where a  model is trained on a fully labeled source home and then used to annotate a new target dataset \cite{dhekane2025transfer}. Thukral et al. \cite{thukral2025layout} achieve this by using text representations of sensor activations as an intermediate modality that allows for a cross-house transfer.

Despite these advancements, the reliance on pre-segmented data remains a significant limitation. Some studies have explored segmentation techniques, such as change-point detection algorithms, which identify abrupt changes in sensor data to infer activity boundaries \cite{aminikhanghahi2018real, sprint2020behavioral, jose2017improving, aminikhanghahi2017using}. However, these methods often rely on heuristics and are not well-suited for capturing fine-grained activities. Moreover, these HAR approaches depend on massive labeled datasets and fixed, pre-defined labels that cannot always capture the nuanced behaviors exhibited in real-world environments. To address this challenge, our work introduces a self-supervised framework to process large unlabeled datasets, paired with clustering and active learning  steps that allow to \textit{discover} fine-grained patterns of daily living with minimal human annotation efforts. 

\subsection{Clustering Pipelines}
\label{subsec:clustering_sh}

Several recent works have explored behavior inference from unlabeled data through automated class discovery. Wu et al.\ \cite{wu2020listenlearner} introduced a pipeline for clustering audio embeddings followed by one-shot labeling, while Patidar et al.\ proposed VAX \cite{vax2023}, which leverages pre-trained video and audio models to bootstrap activity recognition. Similarly, Gao et al.\ \cite{gao2024unsupervised} utilize Large Language Models (LLMs) to iteratively cluster and relabel activities within pre-segmented datasets. 
While these systems demonstrate the utility of user-in-the-loop refinement, their reliance on high-bandwidth modalities like audio and video creates a significant gap in transferability to ambient smart-home environments. Unlike acoustic or vision-based models, which benefit from rich, pre-trained feature extractors (e.g., VGGish or VideoMAE), ambient sensor streams consist of sparse, low-dimensional triggers that lack inherent semantic descriptors. Furthermore, these approaches also require pre-segmentation and assume access to rich multi-sensor pre-training signals.

In the smart home domain, unsupervised clustering methods have been employed in HAR to group unlabeled sensor data into meaningful activity clusters \cite{ariza2020unsupervised}. Classic approaches, such as Lloyd’s clustering algorithm combined with k-Nearest Neighbors \cite{fahad2014activity} and DBSCAN \cite{hoque2012aalo}, have been used to identify patterns in sensor data. More recently, the SCAN framework \cite{scan2020} has been adapted for HAR, enabling the clustering of unlabeled movement data from wearable sensors \cite{ahmed2022clustering}. These methods leverage large-scale sensor data to enhance activity recognition, addressing the challenges associated with obtaining annotated datasets. However, existing ``cluster then classify'' approaches still rely on downstream labels to match and often lack the ability to \textit{discover} fine-grained, data grounded activities.

Recent work by Bock et al.\ \cite{bock2024weak} introduced a clustering-based annotation pipeline that reduces the need for human labeling by leveraging vision-based models for weak annotation in HAR datasets, demonstrating that clustering methods can match fully supervised classifiers while minimizing manual effort. Inspired by successes in the wearable domain, we propose a novel approach tailored for ambient sensor data in smart homes that addresses the challenges of heterogeneous sensor streams while \textit{discovering} fine-grained patterns of daily living rather than merely mapping back to the original labels.

%% file: sections/3_approach.tex
\section{
    DISCOVER:  A Self-Supervised Approach of Identifying Patterns of Daily Living in Human Activities from Smart Home Data
}

\begin{figure}
    \centering
        \includegraphics[width=0.9\linewidth]{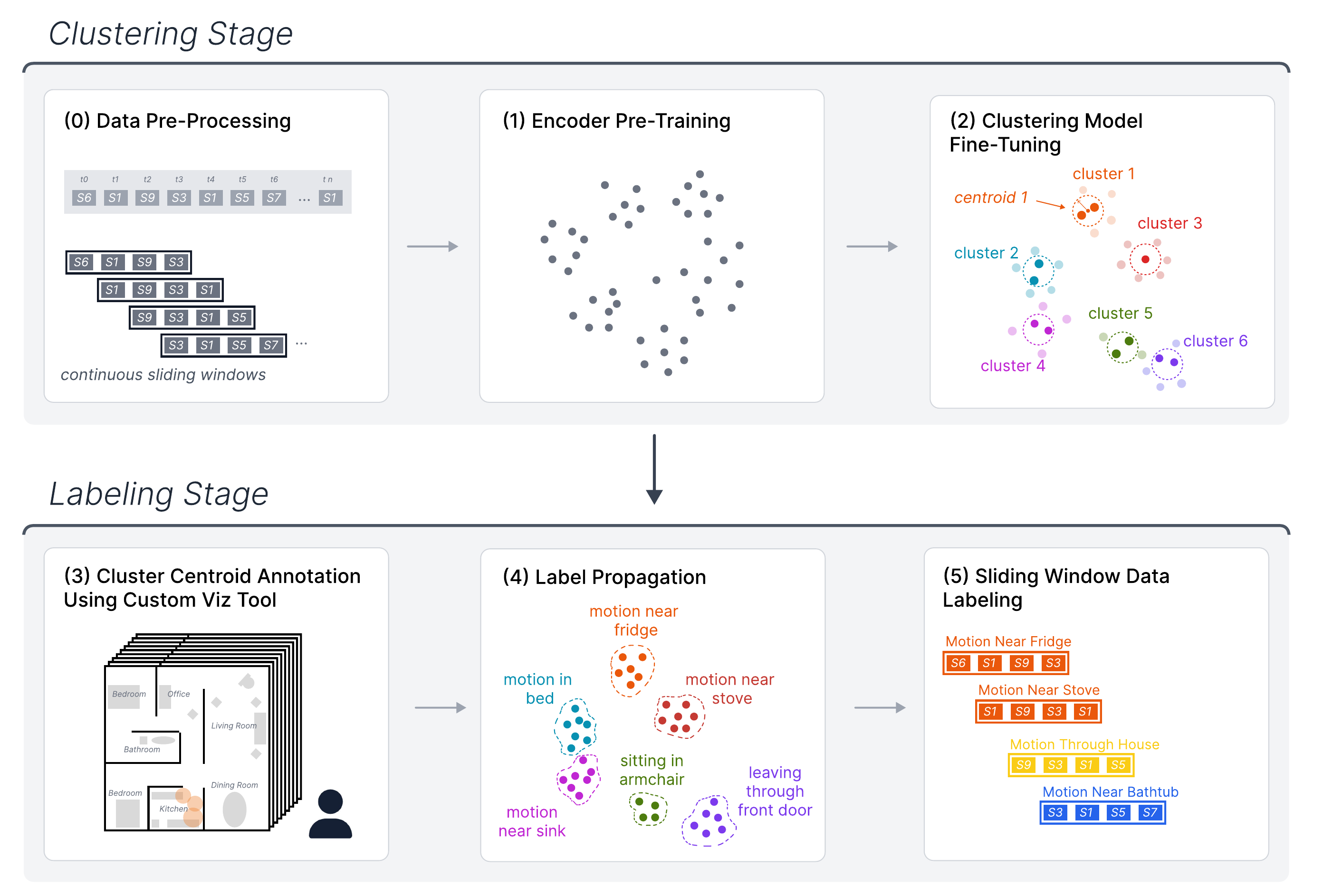}
    \caption{
        Overview of \ToolName{}, a self-supervised framework for the data-grounded discovery of Patterns of Daily Living (PDL), consisting of two main stages: clustering and labeling.
        After (0) slicing the raw data into continuous sliding windows of sensor activations without assuming any pre-segmentation, we procced to the (1) Representation Learning phase, where a Transformer-based model is trained via masked language modeling to map windows into an embedding space, followed by a (2) clustering phase, where we
        use these embeddings to identify similar activity windows and fine-tune a clustering model using SCAN loss, which results in assigning all data points to \( k \) clusters.
        We then (3) sample a handful of windows closest to each cluster centroid, replay them on 2D house layouts using our custom built visualization tool, and send these samples to a group of experts for annotation. With minimal labeling effort, we obtain the annotations of the \textit{discovered} PDLs for each cluster centroid, and (4) propagate them to the rest of the data points in each cluster. These annotations are then (5) applied to the original dataset. The resulting high-resolution PDL stream can be utilized for specialized tasks, such as longitudinal monitoring of behavioral drift or training data-efficient recognition models.
    } 
    \vspace*{-1em}
    \label{fig:main_image}
\end{figure}

To address the limitations of the label-centric paradigm in smart home research, we introduce \ToolName{}, a framework for the data-grounded discovery of \textit{Patterns of Daily Living} (PDL). Unlike traditional approaches that attempt to map sensor streams to a closed set of activity categories, \ToolName{} identifies recurring, fine-grained routines directly from unlabeled data without requiring pre-segmentation. By combining self-supervised representation learning with an interactive refinement loop, the system extracts the idiosyncratic behavioral motifs unique to a resident's specific environment.

We define the objective as the unsupervised discovery of latent behavioral structures in ambient smart home environments. Given a continuous time-series dataset of sensor events, we partition the stream into a collection of $n$ sliding windows $\mathcal{W} = \{W_1, W_2, \dots, W_n\}$ of length $l$ and stride $t$. Our goal is to learn a mapping $f: W_i \mapsto c_k$
where $c_k$ represents a specific \textit{Pattern of Daily Living} (PDL) cluster discovered from the underlying distribution of the data. 
In this formulation, $c_k$ does not denote a predefined semantic category (e.g., ``Cooking''); rather, it represents a data-driven routine (e.g., a specific sequence of movements between a dining area and a kitchen).

\cref{fig:main_image} gives an overview of \ToolName{}, which consists of two main stages: clustering and labeling. The approach starts by pre-processing sensor data into sliding windows in step (0). We then pursue the \textit{discovery} through the following five steps:

\begin{enumerate}
   \item {\textbf{Encoder Pre-Training}: 
        Create an embedding representation of sensor readings by training a transformer model with a self-supervised masked language modeling objective. Transformers have been shown to be effective for modeling long-range structure in event-driven sensor streams \cite{you2025crossmodal}, and we treat sequences of ambient sensor events analogously to sentences. This enables the model to learn context-aware representations that capture temporal dependencies and co-occurrence structure in daily routines. We then use these embeddings to create a set of \( m \) nearest neighbors for each window \( W_i \).
        }

    \item \textbf{Clustering Model Fine-Tuning}: 
        Fine-tune the pre-trained encoder using SCAN loss \cite{scan2020}, which encourages consistency among nearest-neighbor embeddings and partitions all windows \( W_i \) into \( k \) clusters. 
        This serves as our primary clustering approach for discovering data-driven Patterns of Daily Living.

    \item \textbf{Centroid Annotation}: 
        Select a handful of windows \( W_i\) closest to each cluster centroid, and send them to expert annotators to have them assign a label to each sample using \ToolName{} custom built visualization tool.
    
    \item \textbf{Label propagation}: 
        Propagate the centroid labels to every data point in the cluster. 
    
    \item \textbf{Re-Annotation of the Original Time-Series Data}: 
        Apply these cluster labels to the original time-series data.
\end{enumerate}

The following sections describe the technical details of each stage of \ToolName{}, beginning with self-supervised representation learning, followed by clustering-based pattern discovery, and concluding with human-in-the-loop annotation and label propagation.

\subsection{Encoder Pre-Training}
\label{sec:enc_pre_train}

\begin{figure}
    \centering
        \includegraphics[width=0.9\linewidth]{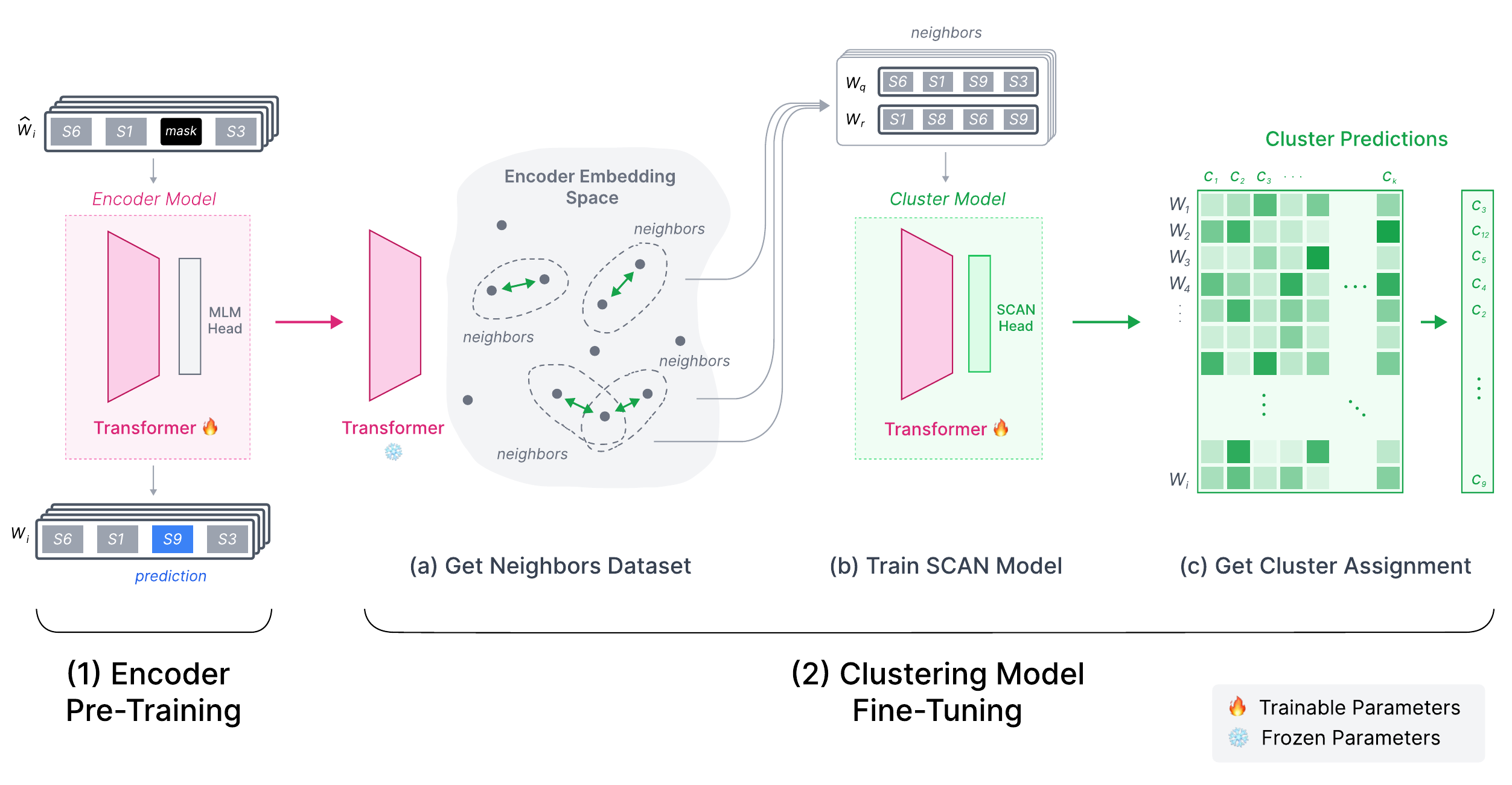}
    \vspace*{-1em}
    \caption{
        \ToolName{} model training pipeline, consisting of (1) Encoder Pre-Training, and (2) Clustering Model Fine-Tuning. 
        \ToolName{} first trains a Transformer model using a Masked Language Modeling head in (1) to obtain semantic embeddings for each window \( W_i \). In step 2(a) it uses these embeddings to identify similar activity windows and pairs them together as a a new training set. It then continues training the fine-tuning the Transformer model with a SCAN loss (2(b)). In the end, the trained SCAN model assigns a cluster \( c_k \) to each input sequence \( W_i \) (2(c)).
    }
    \vspace*{-1em}
    \label{fig:training_pipeline}
\end{figure}

The first step is to learn an informative representation from raw sensor sequences. In this subsection, we describe how we train a Transformer sequence model using a masked language modeling (MLM) objective to capture context-rich representations of ambient sensor data without requiring any labels.

We build on the approach from Hiremath et al. \cite{hiremath2022bootstrapping}, which employs a pre-trained BERT Transformer model to derive embeddings for sensor sequences by feeding in new, domain-specific tokens and training for several epochs.
In addition to the advantages demonstrated in \cite{hiremath2022bootstrapping}, we also note that recent work in the ambient-sensor HAR community has shown that Transformer-based architectures outperform recurrent models such as BiLSTMs. In particular, You et al.\ \cite{you2025crossmodal} demonstrate that Transformers provide better modeling of long-range temporal dependencies, handle sparse event-driven sequences more effectively than recurrent architectures, and offer improved generalization across heterogeneous home layouts by learning richer contextual structure.
Unlike the approach proposed by Hiremath et al. in \cite{hiremath2022bootstrapping}, our pipeline does not rely on pre-segmented data, staying faithful to a more realistic, fully unsupervised scenario. \cref{fig:training_pipeline} (1) shows a diagram of this process. First, we load the original BERT model that was pre-trained on text corpora (BooksCorpus and English Wikipedia) \cite{bert2019} from HuggingFace \cite{huggingface2020} and extend the original vocabulary of size \texttt{30,522} WordPiece tokens by adding sensors and their readings as new tokens to the original vocabulary (which evaluates to roughly \texttt{100} new tokens per dataset). For example, sensor labeled \texttt{M1} and its output signal \texttt{ON} are concatenated into a new token \texttt{M1\_ON}, assigned a new token ID, and initialized with random \texttt{768}-dimensional embedding. Once all training input sequences are processed and encoded, we train BERT model using a self-supervised MLM objective, defined below. 

For every input sequence \( W_i = \{d_1, d_2, \dots, d_l\} \) of sensor tokens \( d \), we choose a proportion \( p \) of these readings to mask. Following \cite{hiremath2020deriving} and the original BERT paper \cite{bert2019}, we set \( p = 0.15 \). Let \( M \subset W_i \) be a subset of positions selected for masking, where \(\lvert M \rvert = \lfloor p \times l \rfloor\). We replace each token in \( M \) with a special \texttt{[MASK]} token, producing a masked sequence \(\hat{W_i}\). The model is then trained to predict the original tokens in \( M \) based on the surrounding context provided by \(\hat{W_i}\).

The training loss \( \mathcal{L}_{MLM} \), which is essentially a cross-entropy over the masked tokens, is defined as:
\[
\mathcal{L}_{MLM} = - \frac{1}{|M|} \sum_{x \in M} \log P(d_x \mid \hat{W_i})
\]

where \( P(d_x \mid \hat{W_i}) \) is the probability that the \(x\)-th masked token \( d \) is correctly predicted by the model.

This loss function incorporates the context from the surrounding, unmasked tokens (i.e., sensor readings) in each sequence. By introducing new tokens for sensor readings (e.g., \texttt{M1\_ON}) into the vocabulary and initializing their embeddings randomly, the model is able to integrate sensor-specific information into the learned representations during pre-training.
Once the training is done, we obtain an embedding for each input sequence \( W_i \) by removing the MLM head and extracting the \texttt{768}-dimensional vector of a special \texttt{[CLS]} token. This \texttt{[CLS]} (``classification'') token is specifically designed to represent the entire sequence in a single embedding, a common practice in BERT-based architectures to obtain a fixed-length representation of any input sequence \cite{bert2019}.

\subsection{Fine-Tuning the Clustering Model}
\label{sec:fine_tune_cluster_model}
Once we obtain an embedding for each window \(W_i\) in the training data, we use these embeddings to measure the similarity between individual sequence vectors, build a ``neighbors'' dataset, and train a clustering model using this data. First, for every input sequence we want to find and store a small set of examples that are most similar to it in the BERT embedding space (\cref{fig:training_pipeline} (2a)). Let \( \text{sim}(W_q, W_r) \) represent the cosine similarity between two embedding vectors \( W_q \) and \( W_r \), defined as:

\[
\text{sim}(W_q, W_r) = \frac{W_q \cdot W_r}{\|W_q\| \|W_r\|}
\]

\noindent For every input sequence \( W_i \), we identify its \( h \)-nearest neighbors using this cosine similarity. Following the original SCAN paper \cite{scan2020}, we set \( h \) equal to 20. These neighbors will be used as training examples in the following step where the SCAN loss will be forcing them to be in the same cluster \( c_k\)

We take the re-trained BERT model from the previous step and fine-tune it by redefining its training objective (\cref{fig:training_pipeline} (2b)). To achieve that, we replace the MLM training head and its cross-entropy loss function with a clustering head paired with the SCAN loss \cite{scan2020}. SCAN loss consists of two components, an instance-level contrastive loss and a cluster-level entropy loss, defined as follows:

\begin{itemize}

    \item 
    \textbf{Instance-Level Contrastive Loss}: Encourages every input sequence and its neighbors to belong to the same cluster. For a sequence \( W_i \), let \( H(W_i) \) denote its set of \( h \)-nearest neighbors. The contrastive loss \( \mathcal{L}_{\text{contrastive}} \) is defined as:

    \[
    \mathcal{L}_{\text{contrastive}} = - \frac{1}{n} \sum_{W_i} \frac{1}{|H(W_i)|} \sum_{W_j \in H(W_i)} \log P(c_i = c_j)
    \]
    
    where \( P(c_i = c_j) \) is the probability that \( W_i \) and \( W_j \) are assigned to the same cluster.

    \item \textbf{Cluster-Level Entropy Loss}: Encourages balanced cluster assignments by ensuring that the distribution of cluster labels is uniform. Let \( P_k \) denote the probability of sequences being assigned to cluster \( k \). The entropy loss \( \mathcal{L}_{\text{entropy}} \) is defined as:
    
    \[
    \mathcal{L}_{\text{entropy}} = - \sum_{k=1}^K P_k \log P_k
    \]

    \item \textbf{Combined SCAN Loss}: The total SCAN loss \( \mathcal{L}_{\text{SCAN}} \) is a sum of the two components:

    \[
    \mathcal{L}_{\text{SCAN}} = \mathcal{L}_{\text{contrastive}} + \lambda \mathcal{L}_{\text{entropy}}
    \]
    
    where \( \lambda \) controls the weight of the entropy loss. Following \cite{scan2020}, we set \( \lambda = 2\).

\end{itemize}

SCAN loss enables the BERT model to assign cluster labels 
\( c_i \in \{1, 2, \dots, k\} \) to each sequence \( W_i \). The clustering head is initialized 
with random labels and iteratively learns to group similar sequences together, thus refining the underlying embeddings into more distinct clusters. Along with the cluster labels (\cref{fig:training_pipeline} (2c)), the trained model provides a probability \( P(c_i = k) \), indicating how confidently each sequence 
\( W_i \) is assigned to cluster \( k \). In the end, every input window receives both a cluster assignment and a corresponding probability distribution over all clusters.

\subsection{Cluster Centroid Annotation}

\begin{figure}
    \centering
        \includegraphics[width=0.8\linewidth]{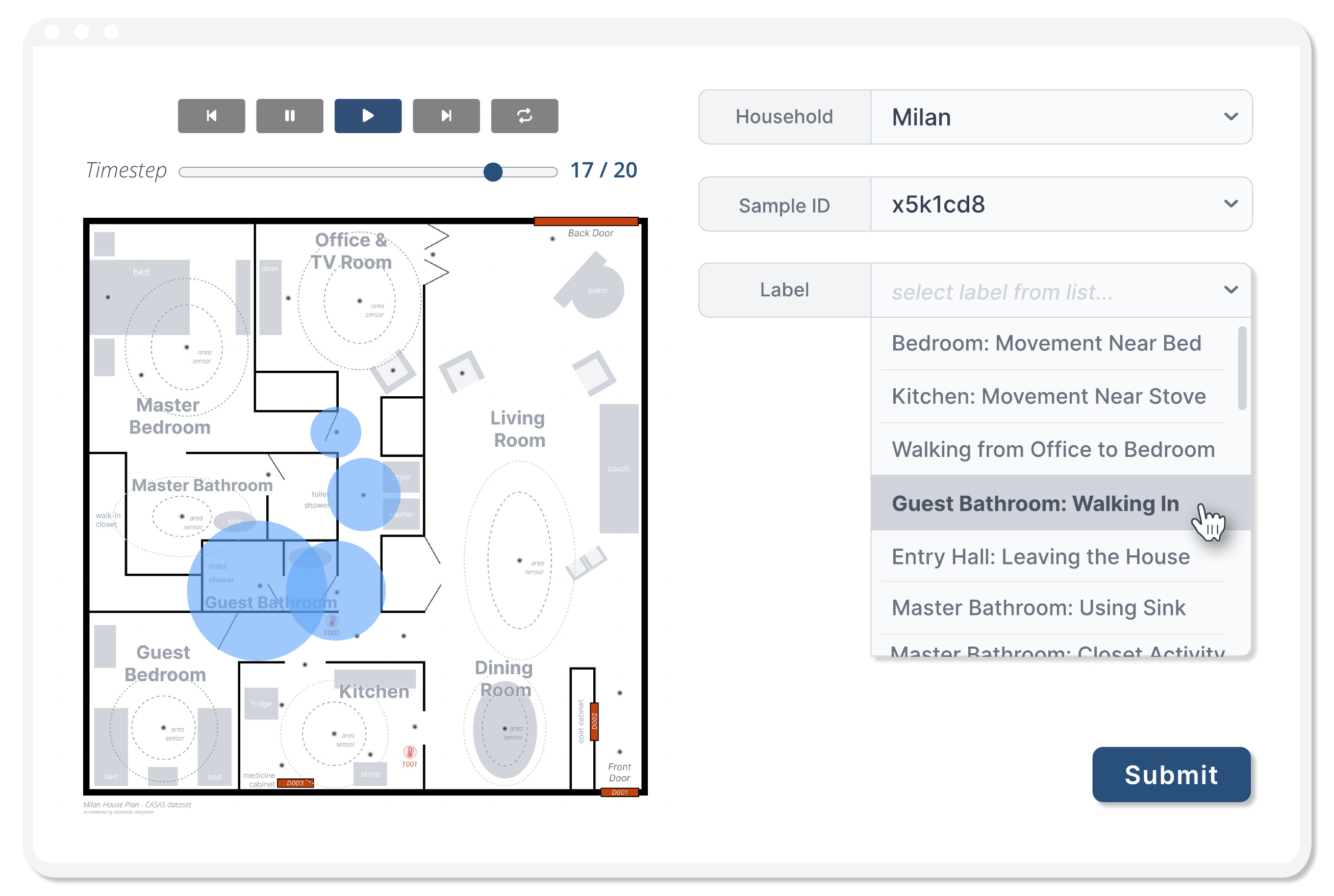}
    \caption{
    \ToolName{} custom-built interactive in-browser annotation tool for reviewing sensor activation sequences \cite{karpekov2025visar} The tool displays a 2D house layout, allowing annotators to replay sequences temporally, observe contextual and spatial details, and assign labels via a drop-down menu. The example above shows a sequence of sensor activations following a resident walking to the guest bathroom. The annotator decides to label this sample as ``Guest Bathroom: Walking In''.
    }
    \label{fig:viz_tool}
\end{figure}

Given the set of clusters learned in the previous stage, the next goal is to interpret and contextualize them by assigning meaningful labels. Since the model has so far operated without any supervision, no ready-made labels exist to inform downstream analyses or practical applications. By annotating a small number of representative samples from each cluster, we can translate the resulting clusters into actionable units—such as identifying specific patterns of daily living or understanding when and where a resident might be performing a certain activity. This process not only validates the clusters internally but also bridges the gap between the automatic discovery of patterns and real-world interpretability.

To minimize the number of sequences we need to review, we will only use \( m \) samples from each cluster that have the highest probability of belonging to that cluster. The probability \( P(c_i = k) \) is taken directly from the SCAN model output. In step (3) in \cref{fig:main_image}, we select \( m \times k \) samples and review them manually using a custom visualization tool, shown in \cref{fig:viz_tool} \cite{karpekov2025visar}. This interactive in-browser annotation tool shows a detailed layout of a given house floor plan and allows to playback the sensor activation sequences temporally, and to assign a label from a drop-down menu. This tool lets the annotator see both the contextual information about that sequence (e.g., location of the activity and its surrounding areas), and the temporal dimension (e.g., whether the activity was static, or if it triggered multiple sensor activations throughout the house). For example, \cref{fig:viz_tool} displays a sequence of sensor activations representing a resident walking through the house to the guest bathroom. The annotator can label this sequence as ``Guest Bathroom: Walking In''. \ToolName{} visualization tool simplifies and speeds up the annotation process and allows for faster iterations.

\subsection{Label Propagation}
\label{sec:label_propagation}

Once each cluster is assigned a label based on its centroid samples, we propagate labels to all remaining sequences belonging to the cluster. This process creates a fully labeled dataset from initially unlabeled sequences. By grouping each sequence with its dominant cluster label, we enable efficient large-scale annotation of resident activities.

\subsection{Re-Annotation of the Original Time-Series Data}
\label{sec:reannotation}

Having assigned labels to each sequence, we then apply these annotations back to the original time-series data, preserving their chronological order. This comprehensive labeling allows researchers to analyze the temporal progression of activities and identify patterns or shifts in behavior over days, weeks, or longer periods. In doing so, our pipeline not only re-annotates large portions of unlabeled data but also lays the groundwork for an array of downstream tasks, such as behavior monitoring, anomaly detection, and personalized interventions.

%% file: sections/4_data.tex
\section{Data Pre-processing and Experimental Setup}
\label{sec:data_prep}

In this section, we provide an overview of the datasets used for our experimental validation and summarize pre-processing steps applied to the data, and describe our labeling tool.

\subsection{Datasets}
\label{sec:datasets}

\input{tables/casas_label_distribution}

Our experimental analysis uses publicly available datasets from the Center of Advanced Studies in Adaptive Systems (CASAS) at the Washington State University \cite{casas2009}. 
The CASAS testbed consists of multiple household datasets, each containing motion (M), door (D), and temperature (T) sensor streams collected over a period of time that stretches multiple months, with associated activity labels. 
In this work, we perform analysis on three households that are widely used in prior HAR work (e.g., \cite{liciotti2020sequential, tdost2024}):

\begin{itemize}
    \item \textit{Aruba:} A single-resident household that spans 220 days and includes M, D, and T sensors.
    
    \item \textit{Milan:} A single-resident household that spans 72 days and includes M, D, and T sensors.
    
    \item \textit{Cairo:} A two-story two resident household that spans 58 dates and contains only M and T sensors.
\end{itemize}

Each CASAS dataset includes assigned activity labels, capturing a range of household activities, such as cooking, relaxing, sleeping. While the labels broadly overlap across datasets, some discrepancies exist: for example, ``guest bathroom activity'' in Milan does not appear in Aruba or Cairo, as there are no guest bathroom sensors in those households. When comparing our approach to CASAS labels, we resolve discrepancies by mapping each dataset's labels to a unified set of standard categories (e.g., \emph{Relax}, \emph{Cook}, \emph{Sleep}), following the method established in prior work \cite{liciotti2020sequential, tdost2024}.

Supporting our motivation for data-driven discovery of patterns of daily living instead of a fixed set of activity labels, we note that a significant proportion of CASAS data has no activity label (\emph{No Label}), ranging from 33\% of data points in Milan, to 53\% in Aruba and 77\% in Cairo. Prior HAR techniques that rely on pre-segmentation excluded this data from their analyses \cite{tdost2024, deepcasas2018, liciotti2020sequential, bouchabou2021fully, bouchabou2021using, you2025cross, gao2024unsupervised, dhekane2025thou}, thereby eliminating 33-77\% of human behavior data from their analysis. As our results will demonstrate, \ToolName{} enables us to cluster and obtain labels for all data, ensuring a more complete analysis of underlying human behavior patterns.

\subsection{Data Pre-Processing}
\label{sec:data_preprocessing}

Following previous research into the optimal window size $l$ for classifying activities in CASAS \cite{krichnan_cook2014}, we select $l = 20$ for all input windows $W_i$. We use sliding windows with a stride of size \texttt{1} to get sequence samples that only differ by one reading. Since using all of this input data would yield an overly redundant training dataset (which can lead to over-fitting \cite{nils_plotz_2015}), we uniformly sample 10\% of these sequences for training. For example, consider the Milan dataset: there are \texttt{433,665} distinct sensor readings in total, which results in \texttt{433,665 - 19 = 433,646} potential input sequences. After taking a 10\% sample, we are left with approximately \texttt{43,364} data points.

To split the data into training and testing sets, we follow the ``leave one day out'' cross-validation strategy: we split the data by random days: 80\% of distinct days of data are kept for training, and 20\% are set aside for testing. By withholding entire days rather than arbitrary sequences, we reduce the risk of data leakage, where the temporal proximity of training and testing sequences might lead to overly optimistic performance estimates. Instead of temporal split where we use the first 80\% of days for training and the last 20\% of days for testing, we chose to select 20\% of days at random. This decision was motivated by the fact that, for some datasets, the final portion of the year would coincide with the U.S. holiday season (late December and early January), during which residents’ activities differ substantially from other periods. By randomly sampling the test set days, we mitigate seasonal biases and avoid severely underestimating the pipeline performance.

\subsection{Defining a Flexible Taxonomy for Pattern Interpretation}

Our final design choice concerns the selection of labels used to interpret the clusters discovered by \ToolName{}, which are used in the drop-down menu in Figure \ref{fig:viz_tool}. While the inclusion of a label set may appear counterintuitive in a pattern-centric framework, it is important to emphasize that \ToolName{} labels are fundamentally different from predefined activity labels commonly used in supervised HAR. Rather than representing a fixed, closed vocabulary of activities that the data must conform to, our labels define a broad space of \emph{possible} behavioral patterns, many of which may not occur in a given dataset. Moreover, sensor data are never pre-labeled with these categories; instead, labels are assigned only after unsupervised pattern discovery, making the approach weakly supervised rather than fully supervised. In practice, \ToolName{} does not require a closed label set at all, labels are introduced here primarily to support systematic analysis and comparison. 

Within this paper, we organize \ToolName{} labels hierarchically to support flexible interpretation at multiple levels of abstraction. At the highest level, patterns are grouped into broad categories of \emph{single-room} and \emph{multi-room} events, with progressively finer-grained subcategories tailored to each environment (e.g., \emph{kitchen} versus \emph{bedroom}). This hierarchical structure allows annotators to select labels at the level of granularity most appropriate for the observed behavior pattern. We include both conventional HAR labels (e.g., \textit{Cook}, \textit{Relax}, \textit{Work}) and as well as provided more detailed options by studying the physical layouts of the CASAS homes and drawing insights from prior work on structural constructs in activities \cite{shruthi_gameofllms}. Examples include activities within a single room (e.g., \textit{movement all over kitchen}, \textit{movement near fridge}, \textit{movement near stove}, and \textit{movement near medicine cabinet}) and movement actions between the rooms (e.g., \textit{walking from bedroom to office}, \textit{leaving guest bathroom}). See the full list of patterns of daily living and their hierarchies in \cref{fig:atomic_activity_tree} in the Appendix.

%% file: tables/casas_label_distribution.tex
\setlength{\dashlinedash}{2pt}   
\setlength{\dashlinegap}{2pt}     
\setlength{\arrayrulewidth}{0.5pt} 

\begin{table}[ht]
    \centering
    \begin{adjustbox}{width=0.5\textwidth,center}
        \begin{tabular}{l|ccc}
            \toprule
            Label & Milan & Aruba & Cairo \\
            \midrule
            Cook             & 29\%   & 17\%   & -    \\
            Eat              & 1\%    & 1\%    & 15\% \\
            Relax            & 16\%   & 22\%   & -    \\
            Work             & 3\%    & 2\%    & 1\%  \\
            Sleep            & 5\%    & 4\%    & 2\%  \\
            Bathing          & 6\%    & -      & -    \\
            Bed\_to\_toilet  & 0.28\% & -      & 0.23\% \\
            Take\_medicine   & 0.13\% & -      & 0.28\% \\
            Leave\_Home      & 1.01\% & 0.11\% & 0.12\% \\
            \cdashline{1-4}
            Other            & 5\%    & 0.03\% & 4\%  \\
            No Label & 33\%   & 53\%   & 77\% \\
            \midrule
            TOTAL \# sensor readings    & 433,665  & 1,716,039 & 726,534 \\
            Total \# Unique days of data & 72       & 220       & 58      \\
            Num labels & 9        & 6         & 6       \\
            \bottomrule
        \end{tabular}
    \end{adjustbox}
    \caption{
        CASAS Datasets \cite{casas2009}: Label Distribution Summary for Milan, Aruba, and Cairo
    }
    \label{tab:casas_label_distribution}
    \vspace*{-1em}
\end{table}

%% file: sections/5_labeling_quality.tex
\section{Results: Reliability and Coherence of \textit{Discovered} PDLs}
\label{sec:label_quality}

In this section, we evaluate the patterns of daily living identified by \ToolName{} and compare our human-labeled clusters to CASAS labels to highlight how our approach provides a more fine-grained and detailed categorization of activities, offering finer resolution and capturing nuances that the original labels may miss. We also evaluate the quality and consistency of the \ToolName{} labels by examining two key metrics: inter-rater agreement, which measures consistency between different labelers, and cluster agreement, which assesses the precision of the labels assigned to each cluster.

\subsection{The Annotation Protocol}

To identify patterns of daily living, we generated $k=20$ clusters per dataset using the steps outlined in Sections \ref{sec:enc_pre_train} and \ref{sec:fine_tune_cluster_model}. To obtain a label for each cluster \( c_k\), we then chose \( m\) sequence samples that are closest to the cluster centroid for labeling, with \( m = 5\) in this work. This resulted in a total of $300$ samples across all three datasets. Samples were randomized and each sample was labeled by two raters, resulting in $600$ total sample-label pairs. We recruited 15 raters, including five graduate students and ten working professionals. Participants had no prior experience with ambient sensor data; thus, their labeling decisions were based solely on the visual representations provided by the \ToolName{} annotation tool \cite{karpekov2025visar}.  \cref{fig:viz_tool} shows a screenshot of a sample annotation replay for the Milan household. During the replay, the rater can view a temporal playback of the sensor activations captured by the data sequence, and use a drop-down menu to select a label, or choose ``Other'' if none of the labels apply. Note that the \ToolName{} tool is fully customizable and is available in open source at \url{https://anonymized}.

\subsection{Evaluation of Consensus and Cluster Homogeneity}

We use two metrics to evaluate the labeling stage.  First, we evaluate inter-rater agreement using Cohen’s Kappa \cite{kappas}.  Higher values represent stronger agreement, with scores over 0.8 representing strong alignment. Observing a high inter-rater agreement on the labeling task would indicate that the data sequences represent consistently interpretable human behavior sequences. Second, we evaluate cluster agreement to examine the uniformity of the \( m\) labels obtained for each cluster.  We use Fleiss’s Kappa \cite{kappas}, which generalizes the idea of Cohen’s Kappa to more than two ratings, to capture the consensus among all 10 labels ($m=5$ by a total of $2$ raters per sample) assigned within a single cluster. A high Fleiss’s Kappa indicates that our raters not only agree between themselves, but also that data sequences captured by the cluster represent a single activity rather than disparate data sequences (i.e., the cluster is internally homogeneous).

Since our labels follow a hierarchical, multi-level taxonomy (see \cref{fig:atomic_activity_tree} in the Appendix), disagreements may sometimes stem from different levels of specificity (e.g., \emph{movement near fridge} vs.\ \emph{movement in kitchen}). To account for this, we also compute Kappa scores at a coarser level by ``leveling up'' each annotated label to its parent category. For instance, if one rater selects \emph{movement near fridge} and the other chooses \emph{movement in kitchen}, these would be treated as the same label at the higher level. Comparing Kappa scores at different levels of the hierarchy helps distinguish minor discrepancies in labeling granularity from dramatically different activity interpretations (e.g., \emph{bathroom activity} vs.\ \emph{reading in armchair}).

\input{tables/kappa_scores}

In this section, we first analyze the inter-rater and cluster-agreement scores, then discuss how the obtained labels differ from CASAS labels.
\cref{tab:kappa_scores} summarizes the inter-rater and cluster-agreement scores for Milan, Aruba, and Cairo. We also refer to \cref{fig:cluster_majority_label_summary} in the Appendix for a detailed table showing the majority label assigned to each cluster and its corresponding vote count.

\subsubsection{Inter-Rater Reliability.}

Inter-rater agreement, measured as Cohen's Kappa, exceeds 0.85 for all three datasets: Milan \(\kappa=0.85\), Aruba \(\kappa=0.89\), Cairo \(\kappa=0.87\). These values indicate very high consistency between the two independent annotators in how they interpreted and labeled the same replay samples. Furthermore, when we ``level up'' each label to its parent category in our hierarchical taxonomy (e.g., merging \emph{movement near bed} into \emph{movement in bedroom}), the Kappa scores rise even further—reaching 0.88 in Milan and over 0.90 in Aruba and Cairo. This gain reflects the fact that any disagreements that did arise between raters were largely in relation to differences in label specificity, such as \emph{sitting in armchair} vs.\ \emph{movement in living room}, rather than fundamentally diverging perceptions of the activity itself.

\subsubsection{Cluster Homogeneity.}

Cluster agreement, measured as Fleiss' Kappa, results in moderately high scores ranging from 0.60--0.63 when annotators used highly specific patterns of daily living labels. Once labels are rolled up to a coarser levels, these scores climb to 0.75--0.81 across the three datasets. 
High values of Fleiss' Kappa highlights cluster homogeneity, demonstrating that discovered clusters correspond to coherent and semantically consistent patterns of daily living rather than arbitrary groupings of sensor windows.
The improvement underscores that most labeling discrepancies within a cluster stem from variation in granularity rather than actual activity disagreements. For instance, in one cluster designated \emph{kitchen activity}, half of the annotators specified \emph{movement near medicine cabinet}, while the other half used \emph{movement in kitchen}. In another cluster, some annotators perceived simultaneous sensor firings in two rooms as \emph{multi-room activity}, while others focused on whichever room had the most events.

\vspace{10pt}
The high rater and cluster agreements confirm that clusters produced by the \ToolName{} pipeline are both interpretable and relatively homogeneous. Moreover, the gains observed when we unify labels at a higher level indicate that our hierarchical taxonomy successfully accounts for the inherent variation in labeling granularity. This can allow researchers to adopt the level of detail most appropriate for their analysis. 

\input{tables/subactivity_mapping}

\subsection{Comparative Analysis: DISCOVER PDLs vs. CASAS Annotations}

In this section, we explore the relationship between the identified patterns of daily living (PDL) and the original CASAS annotations. \cref{tab:subactivity_mapping} illustrates how each discovered PDL label maps onto broader CASAS categories (e.g., \emph{Cook}, \emph{Relax}, \emph{Sleep}), along with the percentage of time-window sequences $W_i$ each label occupies. While the overall distribution of activities is consistent across both protocols, \ToolName{} offers two distinct advantages: 
\emph{i)} finer granularity; and 
\emph{ii)} contextual flexibility. First, \ToolName{} provides a highly specific breakdown of broad categories. For example, in Milan, we distinguish between \emph{sitting in a living room armchair} versus an \emph{office armchair}, and in Aruba, a single \emph{Relax} label is further deconstructed into \textit{sitting on the couch} versus \textit{motion in a TV chair}.
Second, and more importantly, \ToolName{} avoids the semantic rigidity inherent in supervised or transfer learning approaches. In those traditional approaches, a physical behavior (e.g., ``sitting in an office armchair'') is often tied to a fixed label (e.g., \emph{Work}) based on source-home assumptions. In contrast, \ToolName{} remains agnostic to such priors, allowing the same physical location to map to different behaviors based on actual usage: in Milan, the office armchair is associated with \emph{Relaxation}, because that's the best place in house to watch the TV from, while in Aruba, it is associated with \emph{Work}. By solely relying on the underlying data patterns without imposing external definitions, \ToolName{} enables more robust longitudinal trend tracking. This allows researchers to first establish a household's unique habits and subsequently detect meaningful behavioral shifts over time without the bias of predefined activity mappings.

\subsection{Visualizing PDLs using t-SNE}
\label{sec:tsne}

\begin{figure}[t]
    \centering
        \includegraphics[width=0.95\linewidth]{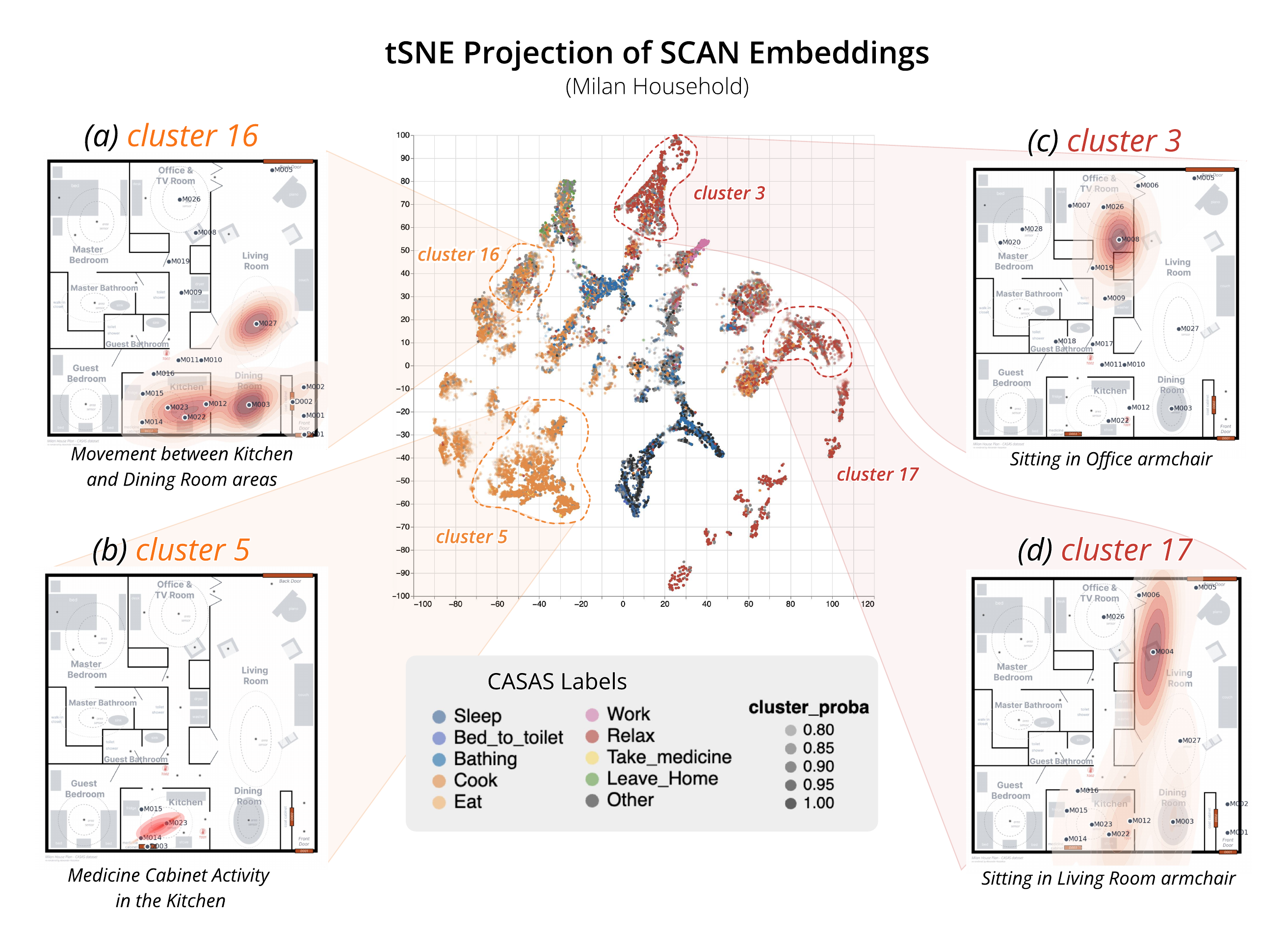}
    \caption{
        tSNE projection of SCAN embeddings from the Milan household; each point represents a sensor window embedding colored by its original CASAS label. Insets display the deployment environment layout overlaid with a heatmap of sensor activations. Insets (a) and (b) highlight two distinct clusters within the CASAS \textit{Cook} label---cluster 16 showing movement between kitchen and dining areas, and cluster 5 capturing activity near the medicine cabinet. Insets (c) and (d) show clusters from the \textit{Relax} label, corresponding to sitting in the TV room armchair and sitting in the living room armchair. Data associated with the Other label (gray) is dispersed across clusters, underscoring its heterogeneous nature. This figure showcases the capability of \ToolName{} to uncover more granular and nuanced behavioral patterns than the original CASAS labels.        
    } 
    \vspace*{-1em}
    \label{fig:cluster_tsne_with_house_maps}
\end{figure}

In this section, we perform a visual inspection of Milan clusters in order to gain more insights into the generated clusters. 
\cref{fig:cluster_tsne_with_house_maps} shows a tSNE \cite{maaten2008visualizing}  projection of the SCAN embeddings from the Milan dataset; each point represents a 768-dimensional sequence embedding colored by its original CASAS annotation. While some activities (e.g., \emph{Work}) form single clusters, data representing other activity labels (e.g., \emph{Cook}, \emph{Relax}) fall into multiple clusters. 
Each point in the scatterplot is annotated with a probability score reflecting SCAN’s confidence in assigning that sequence to its respective cluster, offering interpretability and a level of control that simpler clustering methods (e.g., k-means) lack.

\cref{fig:cluster_tsne_with_house_maps} insets take a closer look at two \emph{Cook} and two \emph{Relax} clusters.
Each inset includes the layout of the deployment environment overlaid with a heatmap representing sensor activation patterns for the data sequences associated with that cluster.  
Insets (a) and (b) are associated with the \textit{Cook} CASAS label (orange).  
Although both sets of data are labeled as \textit{Cook} within CASAS, we observe clear differences between the sensor patterns within each cluster.  
Cluster 16 (inset (a)) captures the person leaving the kitchen and moving to the dining room, with motion sensors from the dining room and part of the dining room registering the person’s presence.  
By comparison, cluster 5 (inset (b)) is tightly focused on activity around the medicine cabinet in the kitchen. 
Insets (c) and (d) are associated with the \textit{Relax} CASAS label (burgundy), where cluster 3 (inset (c)) captures a static activity in the TV room armchair while cluster 17 (inset (d)) corresponds to sitting in a living room armchair instead.  
This visualization illustrates a key advantage of \ToolName{}: its ability to reveal household-specific, fine-grained patterns of daily living, such as sitting in different armchairs or engaging in distinct kitchen routines, that are difficult to capture using predefined activity labels.
In contrast, fully supervised or transfer-based HAR approaches typically collapse such behaviors into coarse categories (e.g., \emph{Relax} or \emph{Cook}), limiting their ability to represent individualized behavioral structure.

%% file: tables/kappa_scores.tex
\begin{table}[ht]
    \centering
    \begin{tabular}{l  cc  cc}
        \toprule
         & \multicolumn{2}{c}{Inter-Rater \( \kappa \) } & \multicolumn{2}{c}{Cluster Agreement \( \kappa \) } \\
        Dataset & Original Label & Level-Up Label & Original Label & Level-Up Label \\
        \toprule
        Milan  & 0.850 & 0.884 & 0.600 & 0.775 \\
        Aruba  & 0.890 & 0.918 & 0.627 & 0.757 \\
        Cairo  & 0.872 & 0.916 & 0.630 & 0.810 \\
        \bottomrule
    \end{tabular}
    \vspace{0.5em}
    \caption{
        Patterns of Daily Living label quality: inter-rater (Cohen’s) and cluster-agreement (Fleiss’s) Kappa scores for each of the three CASAS datasets (Milan, Aruba, and Cairo). ``Original Label'' indicates the use of the most specific labels chosen by each rater (e.g., \emph{movement near fridge}), while ``Level-Up Label'' merges them into coarser categories (e.g., \emph{movement in kitchen}). 
        Overall, very high inter-rater agreement indicates very high alignment between annotators; decent cluster agreement also implies relative homogeneity among the ratings.
        The resulting increase in Kappa scores from ``Original Label'' to ``Level-Up Label'' highlights that most disagreements stem from label specificity rather than fundamentally different interpretations of the underlying activity.        
    }
    \vspace*{-2em}
    \label{tab:kappa_scores}
\end{table}

%% file: tables/subactivity_mapping.tex
\begin{table}[ht]
  \centering
  \begin{adjustbox}{width=\textwidth,center}
    \small
    \begin{tabular}{ll|cc|cc|cc}
      \toprule
      \multicolumn{2}{c|}{} & \multicolumn{2}{c|}{Milan} & \multicolumn{2}{c|}{Aruba} & \multicolumn{2}{c}{Cairo} \\
      CASAS Label & DISCOVER Label & CASAS & DISCOVER & CASAS & DISCOVER & CASAS & DISCOVER \\
      \midrule
      \multirow{5}{*}{Cook} 
          & TOTAL                              & 31\%  & 19\%  & 39\%  & 24\%  & --   & --   \\ 
          \cdashline{2-8}[0.5pt/2pt]
          & Movement all over kitchen          &       & 4\%   &       & 24\%  &       &       \\[0.8em]
          & Movement near Fridge               &       &       & --    & --    & --    &  --   \\[0.8em]
          & Movement near Stove                &       & 2\%   & --    & --    & --    &  --   \\[0.8em]
          & Movement near Medicine Cabinet     &       & 13\%  & --    & --    & --    &  --   \\
      \midrule
      \multirow{4}{*}{Eat} 
          & TOTAL                              & 2\%   & 6\%   & 4\%   & 6\%   & 72\%  & 67\%  \\ \cdashline{2-8}[0.5pt/2pt]
          & Movement all over Dining Room      &       &       &       &       &       &       \\[0.8em]
          & Movement through Kitchen + Dining  &       & 6\%   &       & 6\%   &       & 32\%  \\[0.8em]
          & Movement through Living + Dining   &       &       &       &       &       & 35\%  \\
      \midrule
      \multirow{6}{*}{Relax} 
          & TOTAL                              & 41\%  & 36\%  & 49\%  & 37\%  & 0\%   & 7\%   \\ \cdashline{2-8}[0.5pt/2pt]
          & Motion all over Living room        &       &       &       &  0\%  &       &       \\[0.8em]
          & Motion in TV chair and area        &  --   & --    &       &  6\%  & --    & --    \\[0.8em]
          & Sitting on couch/armchair          &       & 19\%  &       & 29\%  &       & 7\%   \\[0.8em]
          & Sitting in office armchair         &       & 15\%  &       &       & --    & --    \\[0.8em]
          & Movement through Kitchen + Living  &       & 2\%   &       & 2\%   &       &       \\
      \midrule
      \multirow{2}{*}{Work} 
          & TOTAL                              & 3\%   & 4\%   & 4\%   & 7\%   & 12\%  & 0\%   \\ \cdashline{2-8}[0.5pt/2pt]
          & Movement near office computer/desk &       & 4\%   &       & 7\%   & --    & --    \\
      \midrule
      \multirow{3}{*}{Sleep} 
          & TOTAL                              & 7\%   & 8\%   & 3\%   & 3\%   & 12\%  & 7\%   \\ \cdashline{2-8}[0.5pt/2pt]
          & Motion in all over Master Bedroom  &       & 8\%   &       & 0.1\%   &       &       \\[0.8em]
          & In Master Bedroom, Movement near bed &       &       &      & 3\%   &       & 7\%   \\
      \midrule
      \multirow{7}{*}{Bathing} 
          & TOTAL                              & 14\%  & 19\%  & --   & --   & --   & --   \\ \cdashline{2-8}[0.5pt/2pt]
          & Leaving the Guest Bathroom         &       &       & --    & --    & --    & --    \\[0.8em]
          & Movement inside the Guest Bathroom   &       &       & --    & --    & --    & --    \\[0.8em]
          & Entering Master Bathroom           &       & 1\%   & --    & --    & --    & --    \\[0.8em]
          & Leaving Master Bathroom            &       &       & --    & --    & --    & --    \\[0.8em]
          & Movement in Master bathroom        &       & 5\%   & --    & --    & --    & --    \\
      \midrule
      \multirow{1}{*}{Leave\_Home} 
          &  Motion near entrance doors                             & 2\%   & 6\%   & 0.3\% & 6\%   & 1\%   & 5\%   \\[0.8em]
      \midrule
      \multirow{1}{*}{--} 
          & Motion through house               & 0\%   & 4\%   & 0\%   & 14\%  & 3\%   & 13\%  \\
      \midrule
      \midrule
      \multirow{1}{*}{} 
          & \textbf{TOTAL}                              & 100\%   & 100\%   & 100\%   & 100\%  & 100\%   & 100\%  \\  
      \bottomrule
    \end{tabular}
  \end{adjustbox}
  \caption{
    Comparison of broad CASAS labels and granular \ToolName{} patterns across three datasets. Percentages show the distribution of labels; dashes ($-$) indicate activities unsupported by a home's specific sensor topology (e.g., no bathroom sensors in Aruba and Cairo). The results highlight \ToolName{}’s ability to identify finer behavioral distinctions while avoiding the semantic rigidity of fixed supervised categories.
    }

  \vspace*{-1em}  
  \label{tab:subactivity_mapping}
\end{table}

%% file: sections/6_supervised_benchmark.tex
\section{Results: Quantitative Evaluation and Sensitivity Analysis}
\label{sec:results:clustering}

In this section, we ignore the \ToolName{} human annotator labels, and evaluate only the quality of the clusters generated in Steps 0-2 of our approach. Examining the clusters helps provide further insights into the degree to which they match previous hand-annotated labels from CASAS, and how they are able to offer an even finer level of detail than the original annotations did. To validate cluster performance, we use CASAS as a ground truth oracle to temporarily assign a label to each cluster through majority voting. We then perform a direct comparison with fully supervised approaches, and investigate the impact of varying cluster count, examining the trade-off between granularity, label alignment and annotation costs.

\subsection{Performance Alignment with Supervised Baselines}
\label{sec:supervised_results}

\input{tables/clf_comparison}

To evaluate the quality of the discovered clusters, we assess their alignment with established CASAS labels. Since \ToolName{} identifies Patterns of Daily Living (PDL) without prior knowledge of these labels, we utilize a majority-vote mapping to assign a semantic category to each cluster based on its constituent samples. This approach allows us to directly compare our self-supervised framework against state-of-the-art fully supervised pipelines: 
DeepCASAS \cite{liciotti2020sequential},
TDOST \cite{tdost2024}, 
and TreeCNN \cite{cao2020treecnn}. Following previous papers \cite{tdost2024, deepcasas2018, bouchabou2021fully, bouchabou2021using, you2025cross}, we utilize the weighted F1 metric to evaluate classification performance, ensuring our results are directly comparable to established benchmarks.
DeepCASAS is a fully supervised deep learning pipeline that leverages LSTM architecture to capture temporal dependencies in the data.
TDOST is transfer learning pipeline that leverages full supervision on the source dataset. TDOST converts sensor triggers into natural language descriptions using LLMs, which provides the model with contextual information and enables the transfer of a single model across multiple households. 
Both DeepCASAS and TDOST rely on pre-segmented activity windows (for both source and target datasets in case of TDOST).
TreeCNN is the last supervised baseline that leverages tree-structured convolutional neural network to. Similar to \ToolName{}, TreeCNN does not rely on pre-segmentation and therefore serves as a suitable baseline.
While surpassing supervised models is not the primary objective of a \textit{discovery}-focused framework, a close performance alignment signifies a reasonable alignment between our PDLs and CASAS labels.  

For a rigorous comparison, we re-trained DeepCASAS and TDOST on continuous sliding windows ($W_i$) to match the realistic, non-pre-segmented operating conditions of \ToolName{}. As shown in Table \ref{tab:clf_comparison_simplified}, \ToolName{} demonstrates remarkable performance alignment with these baselines. The performance delta, reported in the final row as ``DISCOVER vs Best,'' reveals that the gap is minimal across all households. In the Milan and Aruba datasets, \ToolName{} remains within 3 percentage points (pp) of the best supervised results. Notably, in the Cairo dataset, \ToolName{} outperforms all supervised baselines, achieving an F1 score of 0.85 compared to the 0.82 achieved by DeepCASAS and TDOST.

These results underscore the high fidelity of our self-supervised embeddings. Without the benefit of training labels, \ToolName{} organizes raw sensor triggers into clusters that are not only semantically coherent but also match or exceed the classification accuracy of models requiring exhaustive manual annotation. These findings suggest that the latent structures \textit{discovered} by \ToolName{} are fundamentally aligned with the canonical activities traditionally defined by human annotators, while offering the added benefit of \textit{discovering} the nuanced patterns discussed in \cref{subsec:act_levels}.

\subsection{The Impact of Representation-Aware Clustering}
\label{sec:kmeans_results}

\input{tables/k-means}

In prior sections, we motivated SCAN as our primary clustering model because it fine-tunes the encoder to enforce local embedding consistency among nearest neighbors. 
In this section, we want to compare SCAN clusters to a simpler geometric method such as k-Means applied to learned BERT embeddings.

\cref{tab:kmeans_scan} shows how SCAN F1 scores reported in \cref{sec:supervised_results} compare to k-Means clusters. Across all three CASAS datasets, k-Means performs substantially worse than SCAN on Weighted F1 scores. This behavior is expected: ambient smart-home data is sparse and highly heterogeneous Pure geometric clustering fails to recover rare behaviors and often condenses multiple unrelated patterns of daily living into the same cluster. As a result, K-Means is unable to form coherent and interpretable patterns of daily living in the way SCAN does. By contrast, SCAN’s fine-tuning phase encourages clusters whose neighbors share similar context, leading to significantly more stable and semantically meaningful clusters. Representation-aware objectives such as SCAN’s consistency loss are essential for \textit{discovering} recurring routines in smart-home environments.

\vspace{1em}

\begin{figure}
    \centering
        \includegraphics[width=0.6\linewidth]{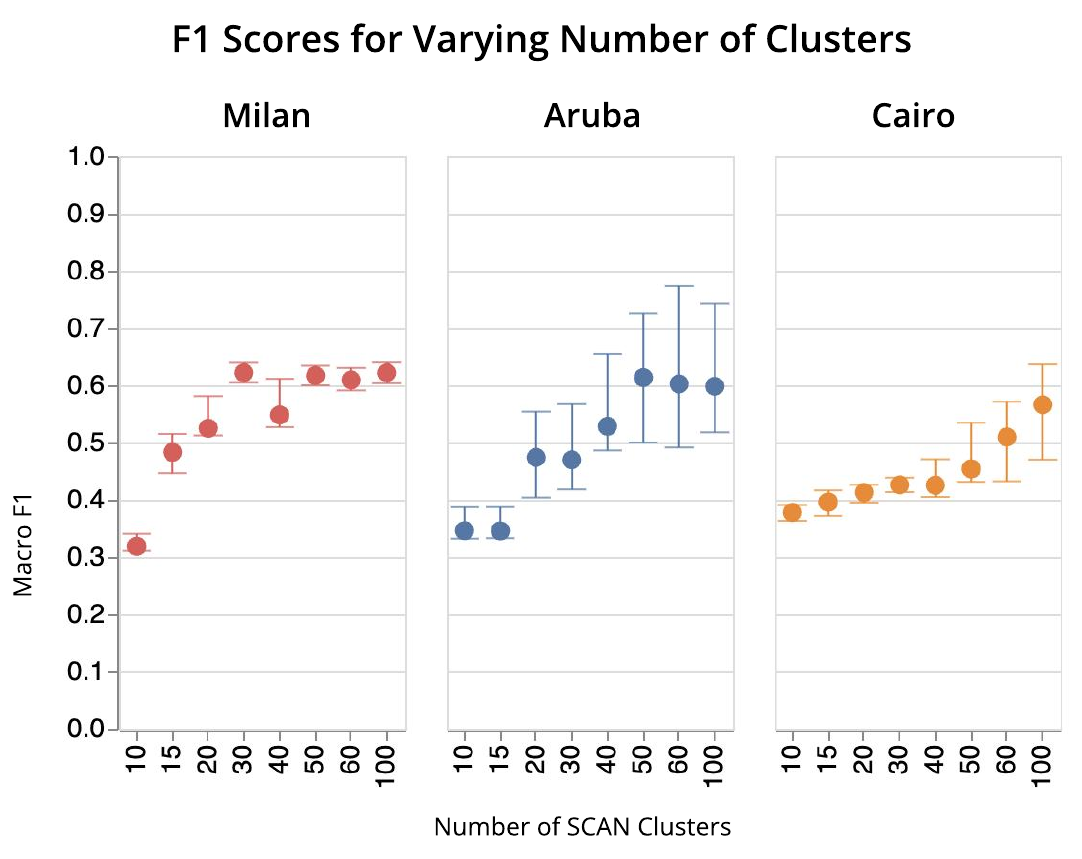}
    \caption{
        F1 scores for varying numbers of SCAN clusters (\(k \)) on Milan, Aruba, and Cairo datasets. The chart shows average F1 scores with bootstrapped 95\% confidence intervals . Increasing \(k\) up until 20-40 clusters improves alignment with CASAS labels before the performance improvement stagnates, suggesting that the optimal number of clusters lies in that range.
    }
    \vspace*{-1em}
    \label{fig:varying_clusters_f1}
\end{figure}

\subsection{Sensitivity Analysis: Cluster Granularity and Labeling Trade-offs}
\label{sec:varying_cluster_size}

We also investigate how the choice of \(k\)—the number of clusters—impacts our model's ability to capture and distinguish fine-grained activities. 
Similar to previous section, we use CASAS labels and apply majority voting per cluster to compute F1 scores. 
For this analysis, we report the Macro-F1 score to ensure equal weighting across all classes. This allows us to assess clustering sensitivity across the entire label distribution, preventing results from being dominated by high-frequency classes and better highlighting the model’s capacity to capture less frequent but clinically relevant activities.

\cref{fig:varying_clusters_f1} shows the Macro F1 scores with bootstrapped 95\% confidence intervals (CIs) for Milan, Aruba, and Cairo, when $k$ varies from 10 to 100. 
The general trend in F1 scores indicates that utilizing more clusters generally leads to better classification performance for all three datasets, although the improvements appear to plateau after \( k = 30\) for Milan, and \( k = 20 \) for Aruba and Cairo. 
This leads to a conclusion that increasing the number of clusters beyond 20–30 is not particularly useful.
\cref{fig:varying_clusters_tsne} in the Appendix shows what these various clusters look like on 2D tSNE projections.

An optimal \( k \) balances classification performance with the workload of labeling cluster centroids. 
We selected \( k=20 \) because it provides a more granular view than the typical CASAS labels while keeping annotation efforts manageable. 

%% file: tables/clf_comparison.tex
\begin{table}[t]
    \centering
    \label{tab:clf_comparison_simplified}
    \small
    \begin{tabular}{llccc}
        \toprule
        \textbf{Model} & \textbf{Method Type} & \textbf{Milan} & \textbf{Aruba} & \textbf{Cairo} \\
        \midrule
        DeepCASAS & Supervised (pre-seg) & 0.79 & \textbf{0.93} & 0.82 \\
        TDOST & Supervised (pre-seg) & \textbf{0.83} & 0.92 & 0.82 \\
        Tree-CNN & Supervised (no pre-seg) & 0.65 & 0.77 & 0.76 \\
        \midrule
        DISCOVER & Self-Supervised (no pre-seg) & 0.82 & 0.90 & \textbf{0.85} \\
        \midrule
        \textbf{DISCOVER vs Best} & --- & -0.01 & -0.03 & +0.03 \\
        \bottomrule
    \end{tabular}
    \vspace{1em}
    \caption{Benchmarking \ToolName{} against SoTA fully-supervised baselines using weighted F1 score. Bold values indicate the best performance for each dataset. The final row reports the absolute gap between the best supervised method and \ToolName{}. \ToolName{} demonstrates near SoTA results (and surpassing Cairo) on non pre-segmented data.}  
    \label{tab:clf_comparison_simplified}
\end{table}

%% file: tables/k-means.tex
\begin{table}[H]
    \centering
    \label{tab:kmeans_scan}
    \small
    \begin{tabular}{lccc}
        \toprule
        \textbf{Method} & \textbf{Milan} & \textbf{Aruba} & \textbf{Cairo} \\
        \midrule
        K-Means & 0.65 & 0.66 & 0.66 \\
        SCAN & \textbf{0.79} & \textbf{0.93} & \textbf{0.82} \\
        \bottomrule
    \end{tabular}
    \vspace{1em}
    \caption{Comparison of K-Means and SCAN clustering performance using F1 score. Bold values indicate the best performance for each dataset. SCAN consistently outperforms K-Means across all datasets, demonstrating the importance of representation-aware fine-tuning.}    
    \label{tab:kmeans_scan}
\end{table}

%% file: sections/7_discussion.tex
\section{Discussion}

In this section, we discuss the real-world potential and practical deployability of \ToolName{} by demonstrating its capacity to transform raw, unlabeled sensor streams into actionable behavioral insights. We first evaluate its ability to extract and monitor data-grounded longitudinal patterns (\cref{sec:discuss_long}), a critical requirement for tracking long-term health trends in unscripted environments. We then highlight the method's utility as an automated annotation engine (\cref{sec:discuss_annotation}), which significantly reduces the human effort typically required to bootstrap new smart home deployments. Furthermore, we discuss how \ToolName{} can re-annotate existing benchmark datasets at finer, more nuanced granularity levels than their native labels currently allow (\cref{sec:discuss_reannotation}), unlocking deeper layers of behavioral analysis. Finally, we outline current limitations and propose future research directions to further enhance the system’s robustness and adaptive capabilities (\cref{sec:limitations_future_work}).

\subsection{Longitudinal Analysis of Behavioral Trends}
\label{sec:discuss_long}

\begin{figure}
    \centering
        \includegraphics[width=1.0\linewidth]{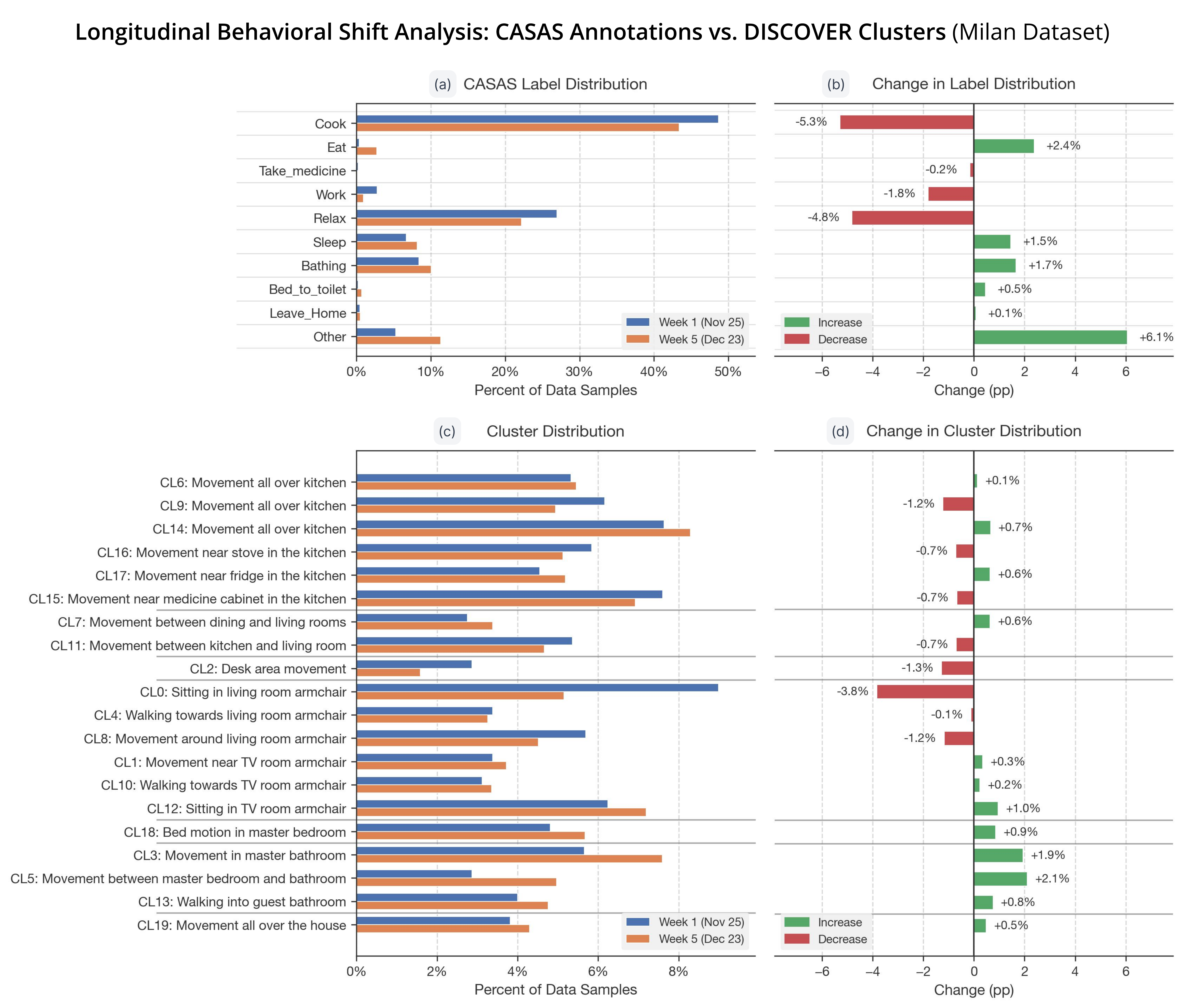}
    \caption{
        Longitudinal behavioral trend analysis on the CASAS Milan dataset. The comparison spans two one-week periods (Nov 25 vs. Dec 23) and highlight shift in routines. Subplots (a) and (b) display macro-shifts in native CASAS annotations, where cooking and relaxing decreased by 4-5 percentage points (pp), while eating, sleeping and bathing increased by 2-3 pp. Subplots (c) and (d) demonstrate that our \textit{discovered} clusters not only recover these high-level trends, such as increased master bedroom and bathroom activity, but provide a more nuanced, data-grounded view. For instance, DISCOVER reveals that the $3.8$ decrease in sitting in living room armchairs (CL0) is partially replaced by sitting in the TV room armchair (CL12), uncovering detailed changes in household routines through an entirely label-free approach.
    }
    \label{fig:longitudinal_trends_usecase}
\end{figure}

The motivation for human activity recognition in smart homes extends beyond the instant detection of actions. 
While traditional HAR systems focus on immediate states, such as sitting, walking, or running, a critical long-term application lies in the longitudinal tracking of behavioral change. 
For example, from a clinical perspective, identifying shifts in daily routines is a cornerstone of health intervention research \cite{dunton2021intensive, varun2019fromsensing}. Longitudinal tracking can validate the success of interventions aimed at reducing sedentary behavior, where success is measured by persistent, month-over-month shifts in activity distribution rather than a single day’s performance \cite{jekel2015mild}. 
Furthermore, in the field of Mild Cognitive Impairment (MCI) and dementia research, observing gradual but persistent changes in habits, such as changes in sleep cycles or increased frequency of nocturnal bathroom visits, can serve as an early indicator of cognitive decline \cite{shahid2024recognizing, attaullah2025dna}. This makes the ability to identify and compare behavioral patterns across different periods a significant contribution to elderly care.

DISCOVER facilitates longitudinal analysis in these domains by offering a data-grounded perspective to track behavioral shifts.
To demonstrate this, we compared the activity distributions of the CASAS Milan dataset across two distinct periods: one week in November and one week in December.\footnote{The choice of these specific dates is dictated by data availability for the Milan dataset; we acknowledge that comparing two weeks during the holiday season is suboptimal for clinical interpretation; however, this exercise serves to validate DISCOVER's ability to quantify longitudinal variance regardless of the specific seasonal or environmental drivers.} By plotting the distribution of our 20 \textit{discovered} clusters against the original CASAS annotations, we can visualize how a resident's behavioral fingerprint evolves. \cref{fig:longitudinal_trends_usecase} illustrates this phenomenon: examining the original CASAS annotations (subplots (a) and (b)), we detected several shifts in how people spent their time, measured by the percentage point (pp) change in the relative number of data samples.
We see a decrease in ``Cooking'' ($-5.3\%$), ``Relaxing'' ($-4.8\%$), and ``Working'' ($-1.8\%$), and an increase in ``Eating'' ($+2.4\%$), ``Sleeping'' ($+1.5\%$), ``Bathing'' ($+1.7\%$), and the catch-all ``Other'' ($+6.1\%$) categories.
DISCOVER recovers similar high-level trends without requiring any ground-truth labels: subplots (c) and (d) clearly show a decrease in some kitchen movements and desk activity and an increase in master bedroom and bathroom activations, but does so without requiring any labels or supervision. More importantly, DISCOVER offers a more nuanced view specific to this household: our pipeline reveals, for example, that sitting in the living room chairs was largely replaced by time in a TV room armchair, and that kitchen shifts were characterized specifically by decreased stove use but increased refrigerator access.
Furthermore, \ToolName{} permits a more nuanced inspection of the non-descriptive ``Other'' category, which exhibited the most significant longitudinal shift in \cref{fig:longitudinal_trends_usecase} with a $6.1\%$ increase. While the original CASAS annotations offer no insight into the nature of this change, \ToolName{}'s comprehensive, label-free mapping captures the entirety of these sensor events, grounding previously obscure shifts in precise, spatially-anchored patterns.

In contrast to the extensively annotated CASAS dataset, the vast majority of real-world smart home environments remain unlabeled. While simpler statistical or rule-based methods may suffice for datasets like \textit{Milan}, which features a sparse deployment of only 30 sensors, these approaches fail to scale as environments grow in complexity. As we introduce denser instrumentation, including contact sensors for cabinets, smart pill dispensers, and energy-load sensors, the ad-hoc configuration required to manually adjust rule-based pipelines for every unique household becomes prohibitive. \ToolName{} overcomes this bottleneck by automatically extracting patterns directly from the available data, bypassing the need for standardized activity definitions. This capability is particularly vital for longitudinal tracking, where the focus shifts toward modeling within-person variability. By comparing a resident's ``old self'' to their ``new self,'' \ToolName{} can detect subtle, practically, and, hopefully, clinically significant behavioral evolutions that are unique to the individual, providing a scalable pathway toward correlating automated sensing with long-term health outcomes.

\subsection{Annotation of Massive Unlabeled Ambient Sensor Datasets}
\label{sec:discuss_annotation}
Smart homes have been studied extensively across academia, industry and medical fields.  While fully annotated datasets such as CASAS \cite{casas2009}, Marble \cite{marble2022}, and Orange4Home \cite{orange2017} are widely used in AI research, they are barely representative of the predominantly unlabeled and noisy data streams typical of most practical smart home environments.
Large-scale medical initiatives such as CART \cite{beattie2020collaborative} (272 homes over 3 years) and TIHM \cite{palermo2023tihm} (56 homes over 50 days) produce orders of magnitude more data, but remain less commonly used due to their lack of activity labels resulting from prohibitively high costs of manual annotation. With \ToolName{}, it becomes feasible to automatically annotate such massive datasets. Users can choose their desired level of granularity by adjusting the number of discovered clusters, and exercise control over the annotation process via an interactive visualization tool that displays sensor activation sequences for each cluster. While such labels may not constitute true ground truth, which is inaccessible without camera data or similar information, our analysis has shown that processing with \ToolName{} results in semantically meaningful activity clusters that can be labeled with high degree of agreement, thus providing insight into common everyday activities. 
We believe that \ToolName{} can expand access to previously unlabeled data, creating opportunities for advancement in the areas such as in-home health monitoring \cite{morita2023health}, early detection of Alzheimer's \cite{alberdi2018smart}, and intelligent support for individuals with dementia \cite{demir2017smart}.

\subsection{Re-annotation of Existing Datasets with Finer Granularity}
\label{sec:discuss_reannotation}
As we have seen in the case of CASAS \cite{cook2012casas}, existing datasets in the HAR domain often provide only coarse labels that capture broad activities, obscuring the rich variability of human behavior. \ToolName{} addresses this shortcoming by enabling the discovery of 	patterns of daily living and variants within the data.
For instance, while a standard dataset might label hours of activity simply as ``Relaxing,'' \ToolName{} allows us to differentiate between distinct patterns of daily living that are otherwise collapsed into a single category. As shown in \cref{fig:cluster_tsne_with_house_maps}, the activity of ``Relaxing'' in this specific household is manifested through two distinct behaviors: sitting in the living room armchair versus sitting in the TV room armchair. While the specific nature of the activity remains unknown, our method grounds these annotations in the physical layout of the home, transforming a vague semantic label into a precise, spatially-anchored markers.
Similarly, in meal preparation, instead of labeling the entire process as \textit{Cooking}, our approach can reveal detailed steps such as time spent at the stove, visits to the pantry, and travel to/from the dining room.  By re-annotating existing datasets with these finer-grained labels, \ToolName{} provides deeper insights into behavioral patterns, which can be particularly valuable in clinically relevant applications and personalized intervention strategies.

\subsection{Future Work}
\label{sec:limitations_future_work}

While our approach demonstrates promising results, it has certain limitations which can be addressed in future work.
First, our current method processes simple sensor activation sequences without considering additional context such as sensor location, time of day, day of the week, or activation duration. Future work could enrich these input sequences by incorporating spatial and temporal contextual information using either special tokens or descriptive embeddings. This integration could facilitate the discovery of more nuanced behavioral patterns, such as differentiating between daytime and nocturnal bathroom visits or distinguishing unique weekday versus weekend routines.

Second, while the current framework successfully identifies Patterns of Daily Living (PDL), the organization of these patterns into a coherent hierarchy still relies on domain-specific manual mapping. Future work should focus on moving beyond fixed label trees toward the automated discovery of data-driven taxonomies. By processing multiple heterogeneous datasets simultaneously, \ToolName{} could potentially learn a universal structural representation of human behavior that transcends individual household variations. Such a transition from discrete pattern recognition to a systematic, hierarchical taxonomy would provide a deeper, data-grounded understanding of the fundamental building blocks of human routines. This evolution would enable the longitudinal analysis of behavior not just as a sequence of activities, but as a dynamic interplay of multi-level behavioral motifs, ultimately allowing for more robust cross-environment generalization.

Finally, while our current approach reduces the volume of data requiring manual annotation, the selection of semantic labels for each discovered cluster still imposes a cognitive burden on the expert. To further streamline this process, future work could explore the alignment of learned sensor embeddings with natural language representations in a shared latent space. Rather than utilizing Large Language Models (LLMs) as black-box annotators (an approach vulnerable to factual hallucinations and a lack of environmental grounding \cite{Huang_2025}), we propose leveraging LLMs to rank and surface candidate descriptors derived from the latent spatial and temporal regularities identified by \ToolName{}. By surfacing the most probable behavioral motifs for a given cluster, we shift the human role from exhaustive classification to an efficient verification process. This data-grounded alignment would maintain high annotation fidelity while significantly reducing the time and expertise required to label complex, longitudinal sensor streams across diverse households.

%% file: sections/8_conclusion.tex
\section{Summary and Conclusion}
\label{sec:conclusion}

In this paper, we introduced \ToolName{}, a framework for discovering Patterns of Daily Living from continuous ambient-sensor streams without any supervision signal. By combining self-supervised Transformer pre-training with representation-aware clustering (SCAN) and a lightweight human-in-the-loop labeling workflow supported by an interactive visualization tool, \ToolName{} \textit{discovers} interpretable, household-specific Patterns of Daily Living (PDL) with minimal annotation effort, providing a data-grounded foundation for longitudinal behavioral analysis. Across three CASAS homes, we show that the resulting clusters are coherent and homogeneous (high inter-rater and cluster agreement), uncover fine-grained structure beyond predefined ADL vocabularies, and achieve classification performance comparable to fully supervised baselines while using orders of magnitude fewer labels. 

Together, these results position \ToolName{} as a robust complement to traditional HAR systems, providing the granular resolution necessary to move beyond broad activity labels. By grounding behavioral patterns in a resident's specific environment, our framework enables the detection of subtle, within-person habitual shifts, that serve as essential digital biomarkers for early-stage cognitive decline and longitudinal health monitoring.

%% file: sections/9_appendix.tex
\appendix
\section{Appendix}

\subsection{tSNE Visualization of SCAN Clusters}

\begin{figure}
    \centering
        \includegraphics[width=0.8\linewidth]{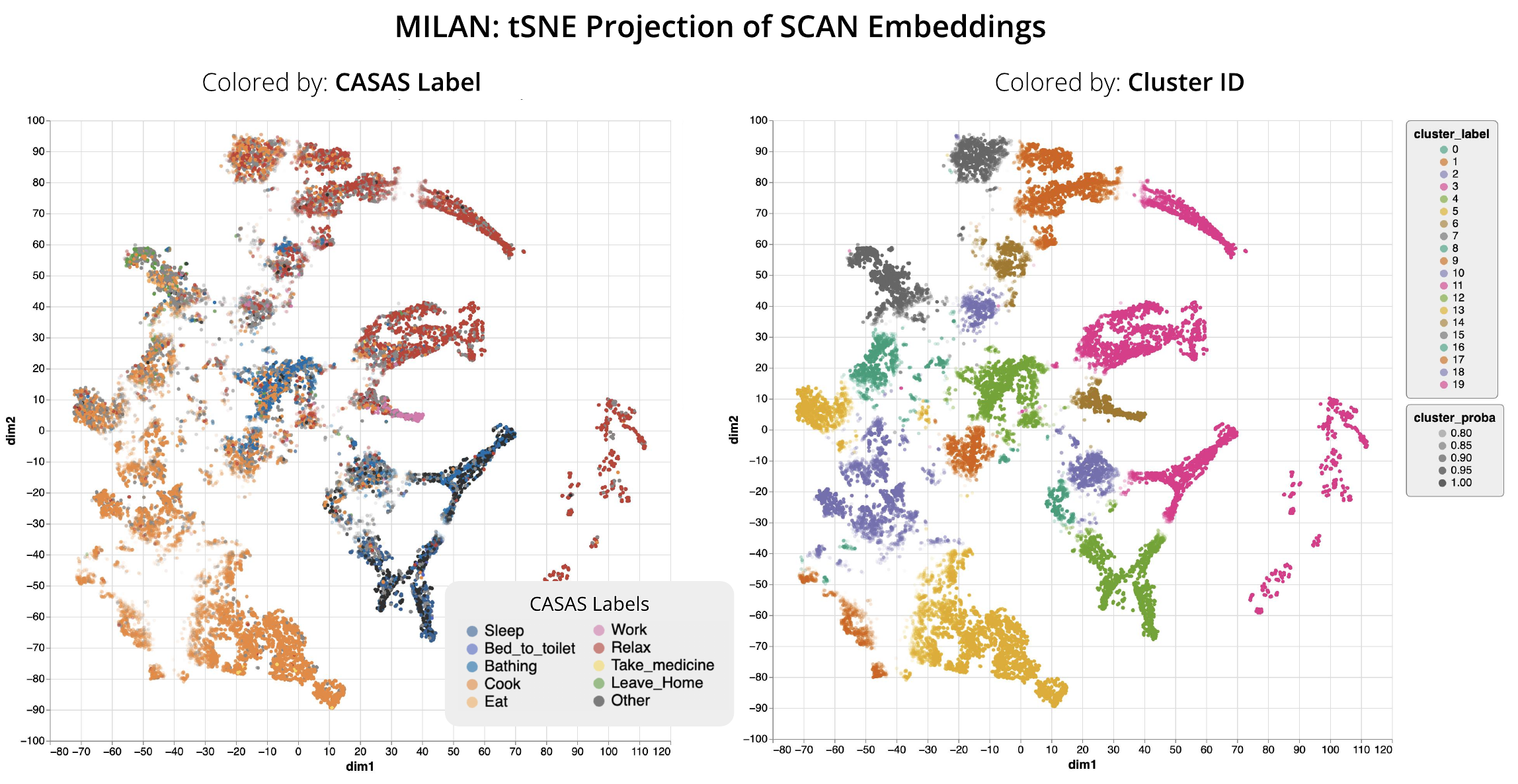}
    \caption{
        tSNE projection of SCAN embeddings from the Milan household. Left pane colored by CASAS labels; right pane colored by SCAN cluster labels from 0 to 19.
    } 
    \label{fig:tsne_label_and_cluster_milan}
\end{figure}

\begin{figure}
    \centering
        \includegraphics[width=0.8\linewidth]{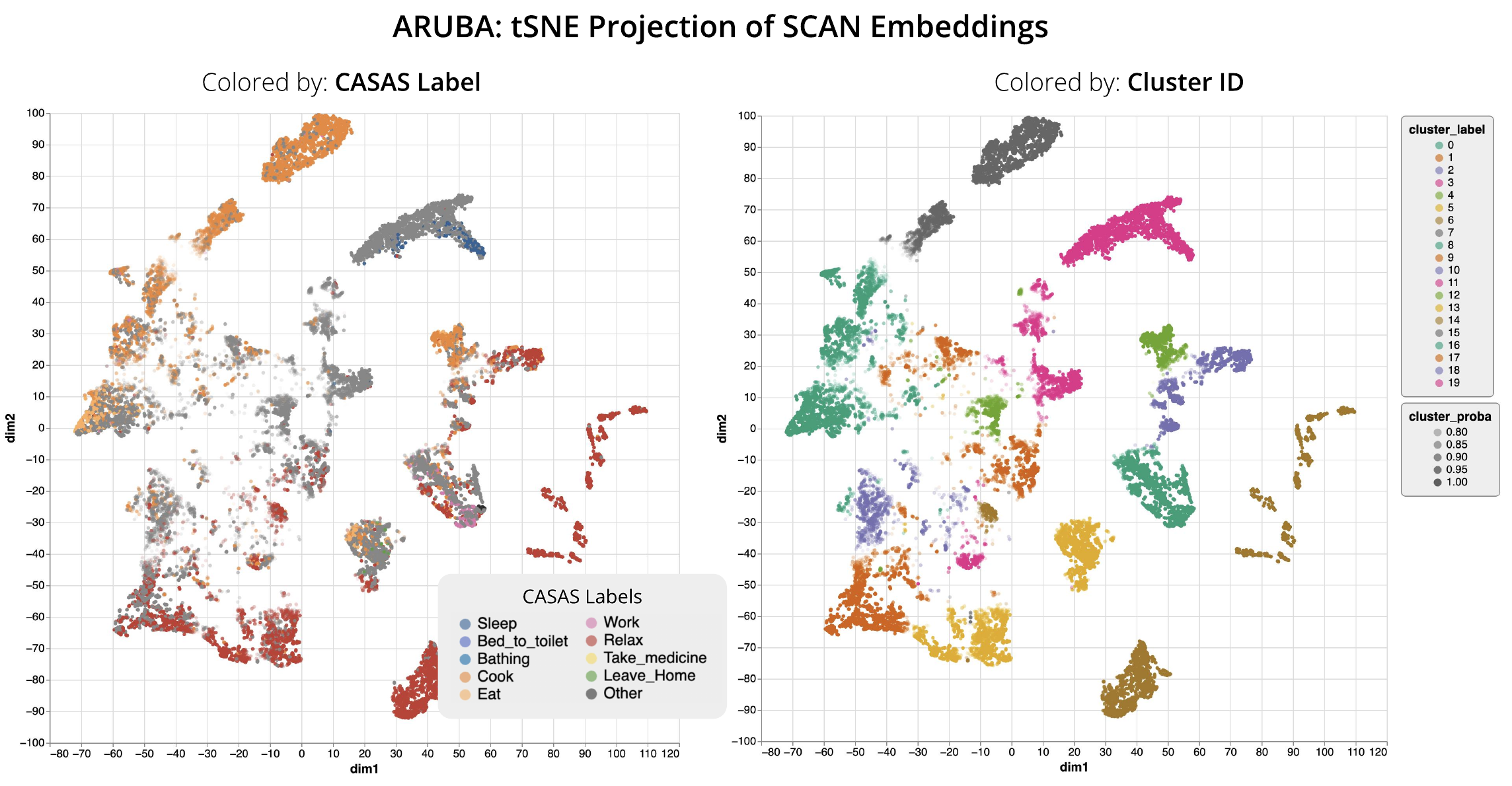}
    \caption{
        tSNE projection of SCAN embeddings from the Aruba household. Left pane colored by CASAS labels; right pane colored by SCAN cluster labels from 0 to 19.
    } 
    \label{fig:tsne_label_and_cluster_aruba}
\end{figure}

\begin{figure}
    \centering
        \includegraphics[width=0.8\linewidth]{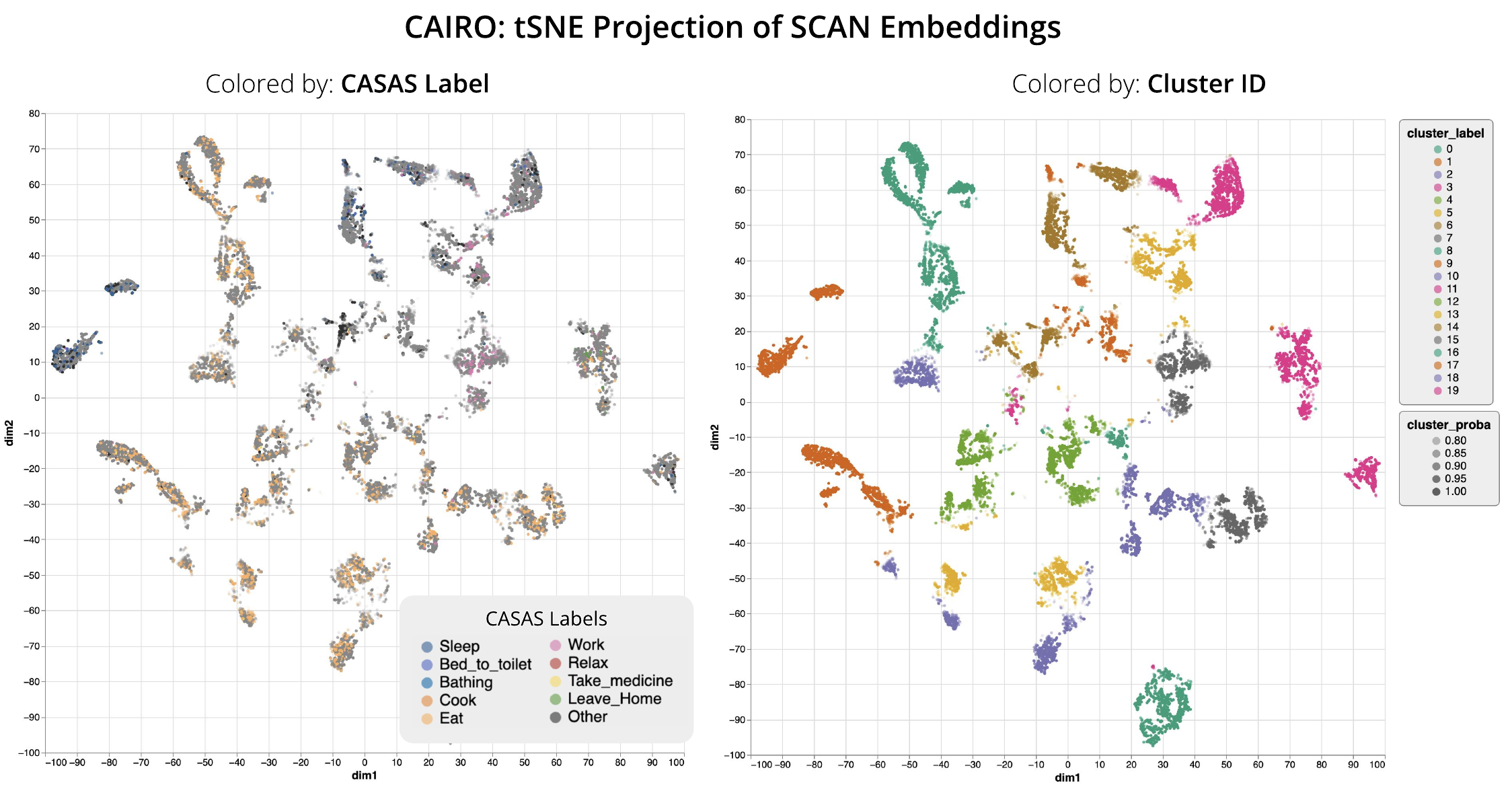}
    \caption{
        tSNE projection of SCAN embeddings from the Cairo household. Left pane colored by CASAS labels; right pane colored by SCAN cluster labels from 0 to 19.
    } 
    \label{fig:tsne_label_and_cluster_cairo}
\end{figure}

Following \cref{sec:tsne}, we present the tSNE projections of \ToolName{} embeddings for all three datasets: Milan on \cref{fig:tsne_label_and_cluster_milan}, Aruba on \cref{fig:tsne_label_and_cluster_aruba}, and Cairo on \cref{fig:tsne_label_and_cluster_cairo}. The left panel shows the embeddings colored by CASAS label; the right one shows the same exact embeddings colored by SCAN cluster labels from 0 to 19. Note how the \textit{Other} label (gray color) is mostly scattered around. Also note that Cairo, which has 81\% of its data points labeled as \textit{Other} (\cref{tab:casas_label_distribution}), has most of its clusters fully gray.

\subsection{tSNE Projection of Varying Number of Clusters}

Following \cref{sec:varying_cluster_size}, we present the tSNE projections of \ToolName{} embeddings for the Milan household for various number of cluster, when \( k \in \{10, 15, 20, 30, 40, 50, 60, 100\} \). The visualization uses CASAS labels to color the clusters, even though they were never used during training. As $k$ increases, the resulting clusters appear to be more distinct and increasingly homogeneous, where each point cloud mostly contains only one distinct CASAS label. As shown in \cref{fig:varying_clusters_f1}, the increase in $k$ is associated with diminishing returns to accuracy, and 20-30 range seems like the best choice for the optimal number of clusters.

\begin{figure}
    \centering
        \includegraphics[width=0.9\linewidth]{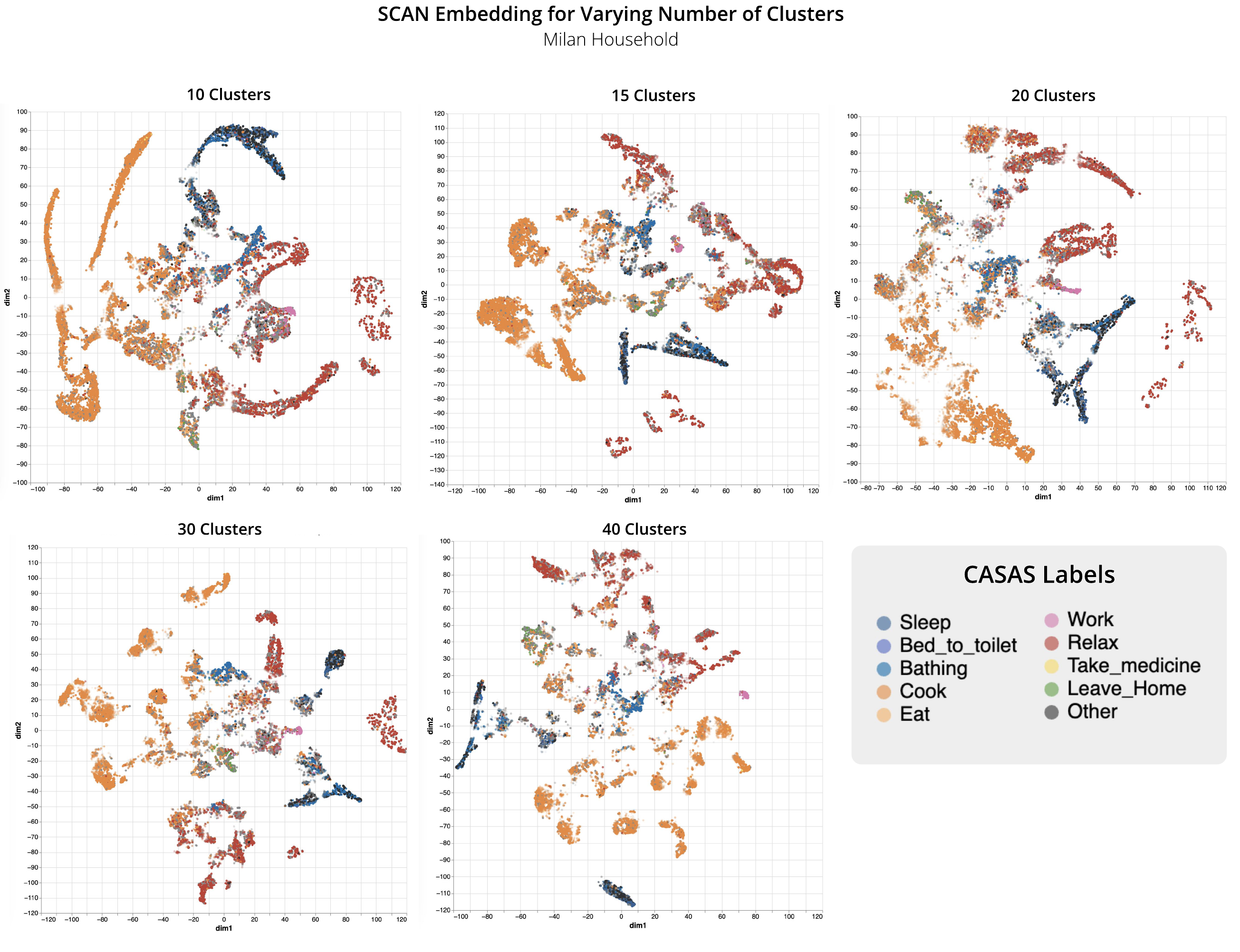}
    \caption{
        tSNE projections of \ToolName{} embeddings for the Milan household with varying numbers of clusters, when \( k \in \{10, 15, 20, 30, 40, 50, 60, 100\} \). Each point represents a sensor sequence colored by CASAS label for visualization purposes only. As \(k\) increases, clusters become more distinct and homogeneous, leading to standalone cluster even for rare activities such as \emph{Leave\_Home}. 
    } 
    \label{fig:varying_clusters_tsne}
\end{figure}

\subsection{Patterns of Daily Living: Labels}

\begin{figure}
    \centering
        \includegraphics[width=0.8\linewidth]{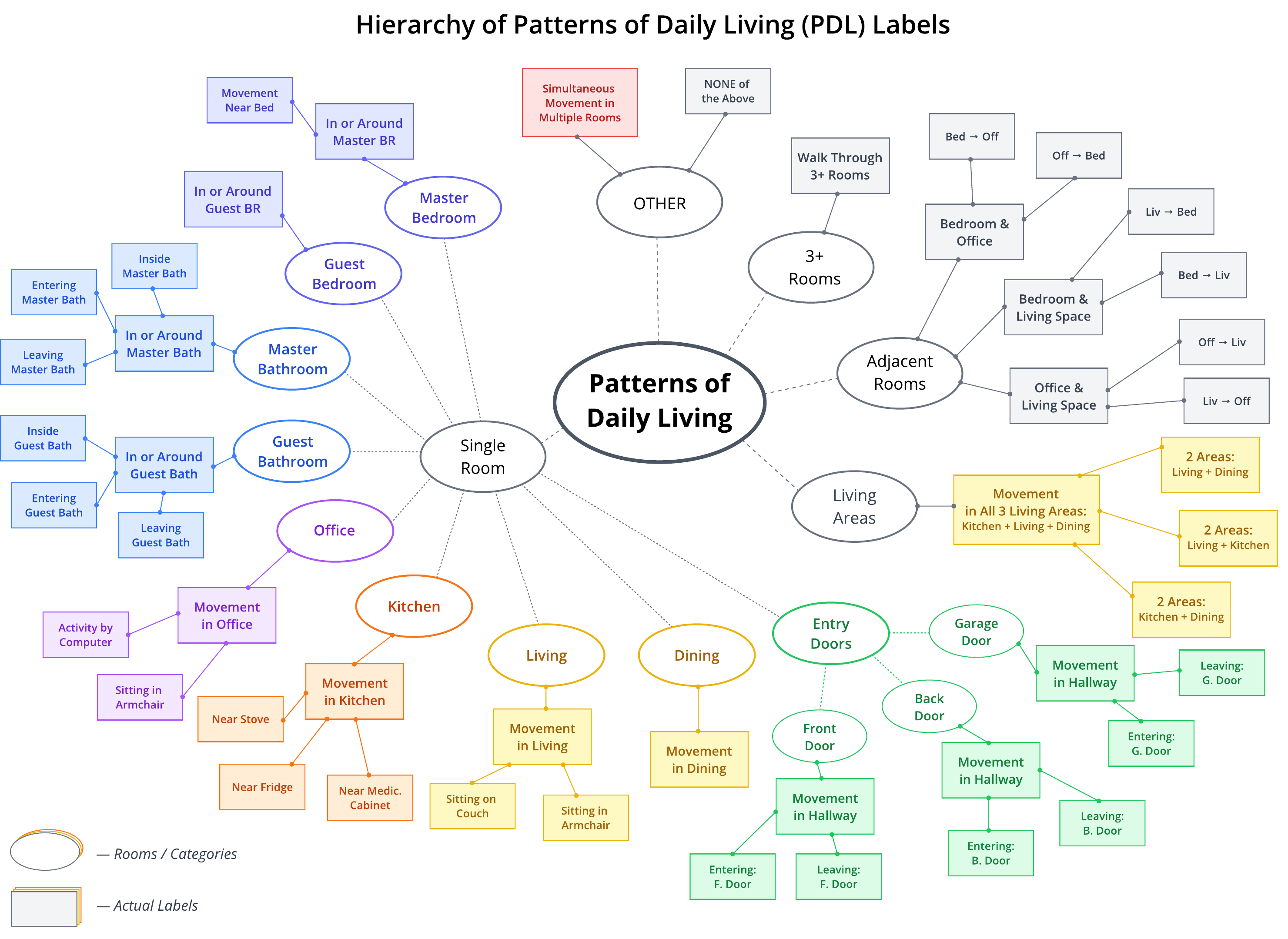}
    \caption{
        \ToolName{} Patterns of Daily Living hierarchy. This list of manually-curated labels was used to populate the labeling drop-down menu in \ToolName{} visualization and labeling tool.
    } 
    \label{fig:atomic_activity_tree}
\end{figure}

\cref{fig:atomic_activity_tree} shows the list of all \ToolName{} PDL labels that we provided the annotators with. The activities are organized hierarchically by categories, e.g. \textit{Single Room Activity} -> \textit{Kitchen Activity} -> \textit{Movement in Kitchen} -> \textit{Movement Near Stove}. These labels were used to populate the labeling drop-down menu in \ToolName{} visualization tool shown in \cref{fig:viz_tool}.

\subsection{Individual Clusters}

\begin{figure}
    \centering
        \includegraphics[width=0.9\linewidth]{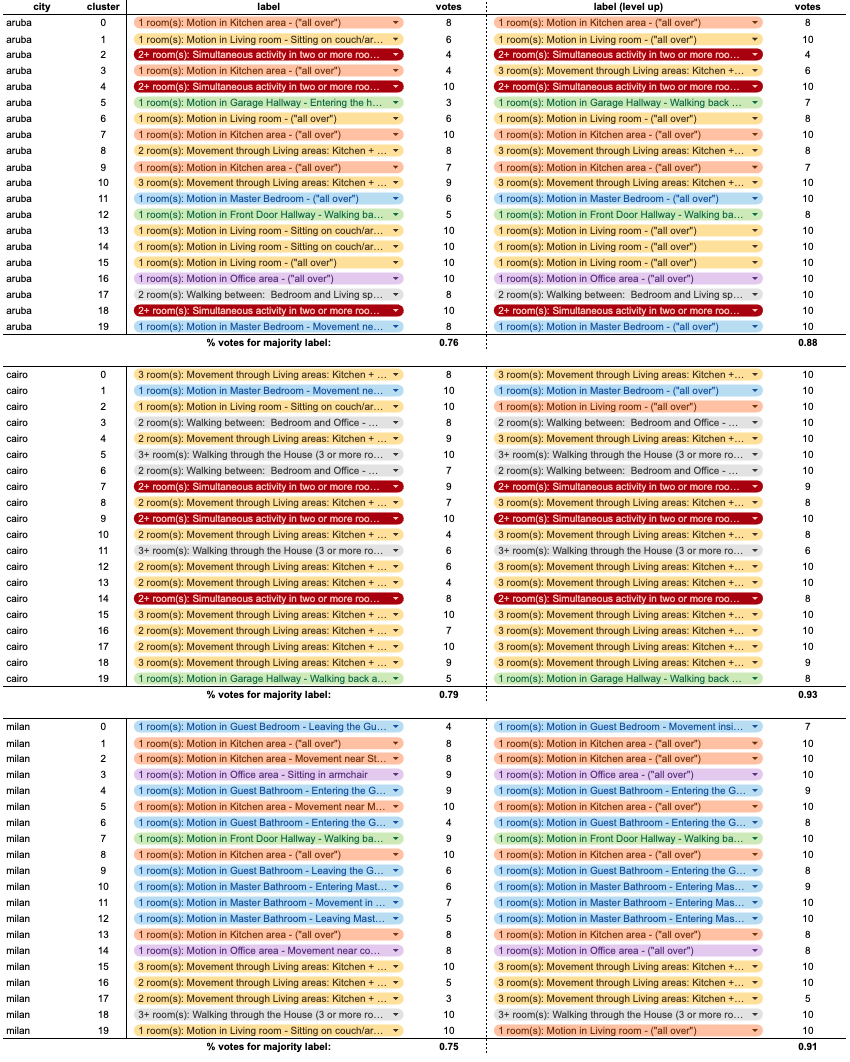}
    \caption{
        	Patterns of Daily Living labels assigned to each cluster  manually labeled clusters for all households. \texttt{votes} column shows how many annotations aligned with this particular label; \texttt{level up} column provides a parent label.} 
    \label{fig:cluster_majority_label_summary}
\end{figure}

Here we present the list of individual cluster labels for all three CASAS datasets. For every cluster, \cref{fig:cluster_majority_label_summary} shows what was the majority label assigned by the annotators, and how many raters "voted" for that label in the \texttt{votes} column. The \textit{label (level up)} column shows up-leveled labels following the hierarchy in \cref{fig:atomic_activity_tree}. This summary provides a closer look at individual cluster annotations that were used to compute Fleiss' Kappa for cluster agreement in \cref{tab:kappa_scores}.

\subsection{Household Layouts}
\label{sec:household_layouts}

\begin{figure}
    \centering
    \includegraphics[width=.95\linewidth]{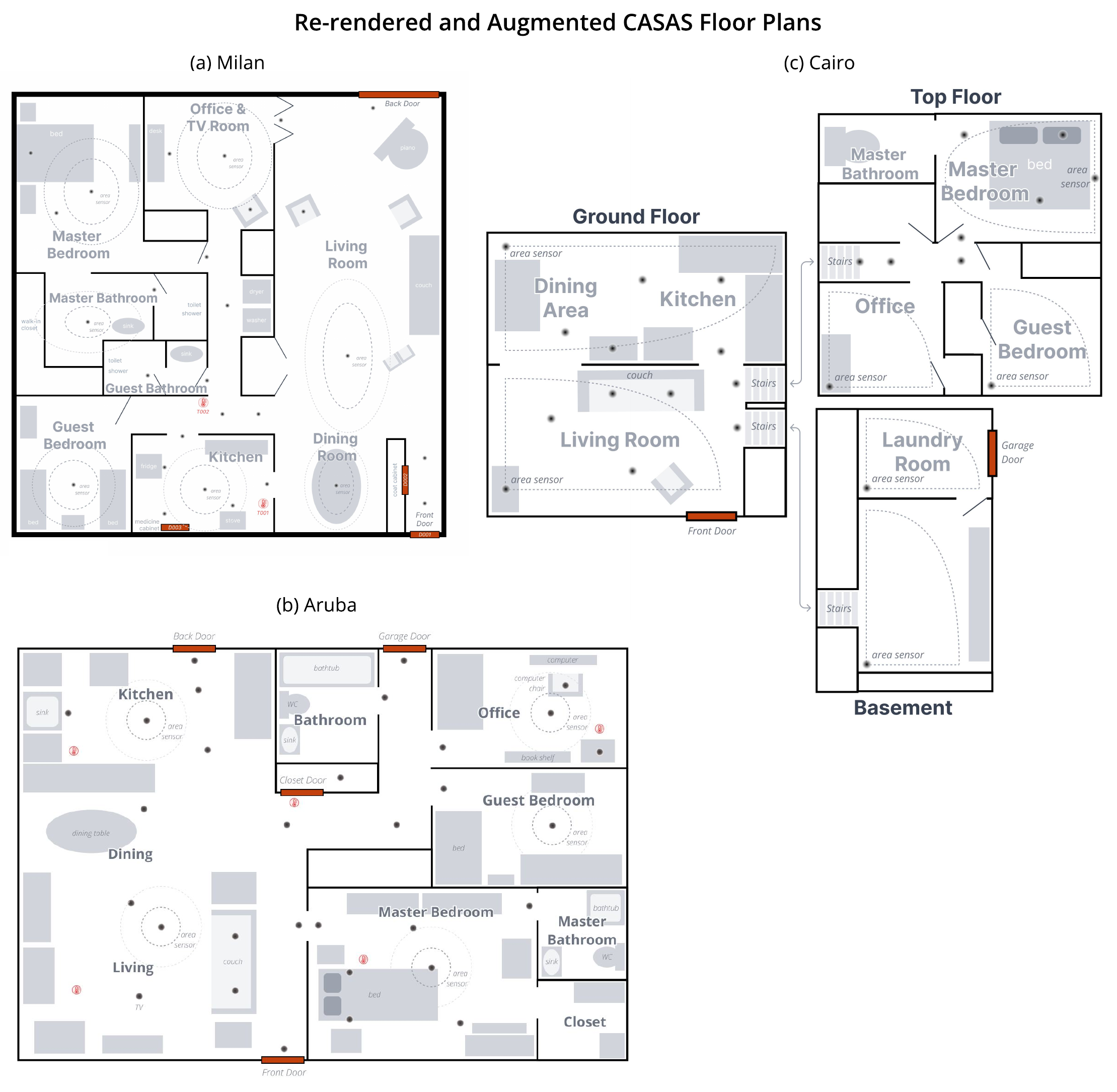}
    \caption{
        Simplified and augmented layout of (a) Milan, (b) Aruba, and (c) Cairo household floor plans, adapted from \cite{casas2009} and \cite{hiremath2022bootstrapping}.
    }
    \label{fig:all_house_layouts}
\end{figure}

We re-rendered the original CASAS \cite{casas2009} household layouts to simplify them visually and make certain household features like furniture stand out more prominently, following the insights from \cite{hiremath2022bootstrapping}.  ~\cref{fig:all_house_layouts} shows the simplified layouts for Milan, Aruba, and Cairo.

